\documentclass[sigconf]{acmart}

\AtBeginDocument{%
  \providecommand\BibTeX{{%
    \normalfont B\kern-0.5em{\scshape i\kern-0.25em b}\kern-0.8em\TeX}}}

\def\FIGDIR{./figs}          

\usepackage{setspace}
\usepackage{comment}
\usepackage[most]{tcolorbox}
\usepackage{pifont}

\usepackage{endnotes,xspace,graphicx,fancyvrb,multirow}

\usepackage{paralist}

\usepackage{mathptmx} 

\newcommand{\ignore}[1]{}
\usepackage{fancyhdr}
\usepackage[normalem]{ulem}
\usepackage{microtype}
\usepackage{flushend}
\newcommand{\PCignore}[1]{}

\newcommand{\insertFigure}[2]{
    \begin{figure}[t]
\setlength{\abovecaptionskip}{0pt}
\setlength{\belowcaptionskip}{0pt}
        \centering
        \includegraphics[width=\linewidth]{\FIGDIR/#1.pdf}
        \caption{\small #2}
        \label{fig:#1}
    \end{figure}
}

\newcommand{\insertWideFigure}[2]{
    \begin{figure*}[h]
        \centering
        \includegraphics[width=\textwidth]{\FIGDIR/#1.pdf}
        \caption{\small #2}
        \label{fig:#1}
    \end{figure*}
}

\newcommand{\insertWideTableFigure}[2]{
    \begin{table*}
    \setlength{\abovecaptionskip}{0pt}
    \setlength{\belowcaptionskip}{0pt}
      \caption{\small #2}
      \label{table:#1}
      \includegraphics[width=\linewidth]{figs/#1.pdf}
      \end{table*}
}

\newcommand{\insertTableFigure}[2]{
  \begin{table}
    \setlength{\abovecaptionskip}{0pt}
    \setlength{\belowcaptionskip}{0pt}
    \caption{\small #2}
    \label{table:#1}
    \includegraphics[width=\linewidth]{\FIGDIR/#1.pdf}
  \end{table}
}

\newcommand{\squishlist}{
 \begin{list}{$\bullet$}
  { \setlength{\itemsep}{0pt}
     \setlength{\parsep}{3pt}
     \setlength{\topsep}{3pt}
     \setlength{\partopsep}{0pt}
     \setlength{\leftmargin}{1.5em}
     \setlength{\labelwidth}{1em}
     \setlength{\labelsep}{0.5em} } }

\newcommand{\squishlisttwo}{
 \begin{list}{$\bullet$}
  { \setlength{\itemsep}{0pt}
     \setlength{\parsep}{0pt}
    \setlength{\topsep}{0pt}
    \setlength{\partopsep}{0pt}
    \setlength{\leftmargin}{2em}
    \setlength{\labelwidth}{1.5em}
    \setlength{\labelsep}{0.5em} } }

\newcommand{\squishend}{
  \end{list}  }

\newcommand{\betterparagraph}[1]{\noindent
\textbf{#1.}}

\newcommand{\colonparagraph}[1]{\noindent\textbf{#1:}}

\newcommand{\TODO}[1]{\textcolor{red}{TODO: #1}}

\newcommand{\tool}{\textsc{MAESTRO}\xspace}

\newcommand{\sMap}{\textsc{SpatialMap}\xspace}

\newcommand{\Cluster}{\textsc{Cluster}\xspace}

\newcommand{\dsetool}{\textsc{COMPOSITORE}\xspace}

\renewcommand{\textrightarrow}{$\rightarrow$}

\begin{document}

\title{Understanding Reuse, Performance, and Hardware Cost of DNN Dataflows: A Data-Centric Approach Using MAESTRO}

\author{Hyoukjun Kwon}
\orcid{0000-0001-9824-1352}
\affiliation{%
  \institution{Georgia Institute of Technology}
  \streetaddress{266 Ferst Drive NW}
  \city{Atlanta}
  \state{Georgia}
  \postcode{30313}
}
\email{hyoukjun@gatech.edu}

\author{Prasanth Chatarasi}
\affiliation{%
  \institution{Georgia Institute of Technology}
  \streetaddress{266 Ferst Drive NW}
  \city{Atlanta}
  \state{Georgia}
  \postcode{30313}
}
\email{cprasanth@gatech.edu}

\author{Michael Pellauer}
\affiliation{%
  \institution{NVIDIA}
  \streetaddress{2 Technology Park Dr}
  \city{Westford}
  \state{Massachusetts}
  \postcode{01886}
}
\email{mpellauer@nvidia.com}

\author{Angshuman Parashar}
\affiliation{%
  \institution{NVIDIA}
  \streetaddress{2 Technology Park Dr}
  \city{Westford}
  \state{Massachusetts}
  \postcode{01886}
}
\email{aparashar@nvidia.com}

\author{Vivek Sarkar}
\affiliation{%
  \institution{Georgia Institute of Technology}
  \streetaddress{266 Ferst Drive NW}
  \city{Atlanta}
  \state{Georgia}
  \postcode{30313}
}
\email{vsarkar@gatech.edu}

\author{Tushar Krishna}
\affiliation{%
  \institution{Georgia Institute of Technology}
  \streetaddress{266 Ferst Drive NW}
  \city{Atlanta}
  \state{Georgia}
  \postcode{30313}
}
\email{tushar@ece.gatech.edu}

\begin{abstract}

%

The data partitioning and scheduling strategies used by DNN accelerators to leverage reuse and perform staging are known as \emph{dataflow}, which directly impacts the performance and energy efficiency of DNN accelerators.
An accelerator microarchitecture dictates the dataflow(s) that can be employed to execute layers in a DNN. 
Selecting a dataflow for a layer can have a large impact on utilization and energy efficiency, but there is a lack of understanding on the choices and consequences of dataflows, and of tools and methodologies to help architects explore the co-optimization design space.
%

In this work, we first introduce a set of data-centric directives to concisely specify the DNN dataflow space in a compiler-friendly form. We then show how these directives can be analyzed to infer various forms of reuse and to exploit them using hardware capabilities. We codify this analysis into
%
an analytical cost model, \tool (Modeling Accelerator Efficiency via Spatio-Temporal Reuse and Occupancy), that estimates various cost-benefit tradeoffs of a dataflow including execution time and energy efficiency for a DNN model and hardware configuration.
%
We demonstrate the use of \tool to drive a hardware design space exploration
%
experiment, which searches across
480M designs
to identify 2.5M valid designs at an average rate of 0.17M designs per second,
including Pareto-optimal throughput- and energy-optimized design points.
%
\end{abstract}

\copyrightyear{2019} 
\acmYear{2019} 
\acmConference[MICRO-52]{The 52nd Annual IEEE/ACM International Symposium on Microarchitecture}{October 12--16, 2019}{Columbus, OH, USA}
\acmBooktitle{The 52nd Annual IEEE/ACM International Symposium on Microarchitecture (MICRO-52), October 12--16, 2019, Columbus, OH, USA}
\acmPrice{15.00}
\acmDOI{10.1145/3352460.3358252}
\acmISBN{978-1-4503-6938-1/19/10}



\maketitle




\section{Introduction}
\label{sec:intro}


Deep neural networks (DNNs) are being deployed at an increasing
scale---across the cloud and IoT platforms---to solve complex regression and classification problems in image
recognition~\cite{simonyan2014very}, speech recognition~\cite{amodei2015deep}, language translation~\cite{wu2016google}, and many more fields, with accuracy close
to and even surpassing that of humans~\cite{karpathy2015deep,toshev2014deeppose,farabet2013learning}.
Tight latency, throughput, and energy constraints when running DNNs have led to a meteoric increase in hardware accelerators.


DNN accelerators achieve high performance by exploiting parallelism over hundreds of processing elements (PEs) and high energy efficiency by maximizing data reuse within PEs and on-chip scratchpads~\cite{eyeriss_isca,chen2014diannao, nvdla,parashar2017scnn,sharma2016high,jouppi2017datacenter}.
For a specific DNN workload and a hardware accelerator, the achieved utilization and data-reuse directly depends on 
(1) how we schedule the DNN computations (e.g., choice of loop transformations) and (2) how we map computations across PEs.
These two components are collectively referred to as \textit{dataflow} in the accelerator literature~\cite{eyeriss_isca, parashar2017scnn, lu2017flexflow, kwon2018maeri}. It has been shown that the energy cost of moving data exceeds the cost of computation~\cite{eyeriss_isca, gao2017tetris}, and so understanding and optimizing dataflow is a critical component of DNN accelerator design, as it directly determines how data is transferred between multipliers (L0), staged in local buffers (L1), and in the global buffer hierarchy (L2 and beyond). 




The performance and energy efficiency of DNN accelerators depend on (1) target DNN model and its layers types/dimensions, (2) dataflow, and (3) available hardware resources and their connectivity.
These three dimensions are tightly coupled, and optimizing DNN accelerators across these dimensions is a challenging task.
For example, a dataflow that exploits input channel parallelism~\cite{nvdla} in convolutional neural networks (CNNs) may not achieve high utilization on layers with a small number of channels.
Alternatively, dataflows that require more transfer bandwidth than the network-on-chip (NoC) provides may result in under-utilization of the hardware. 
In such cases, increasing the L1 scratchpad size may allow the same dataflow to require less data bandwidth, but this larger L1 may increase area and energy consumption. 
%
%
%
Thus, co-optimizing the hardware microarchitecture and the dataflows it supports is 
 one of the primary optimization targets for any accelerator design.
 This remains an open challenge, as observed by the number of novel dataflows and microarchitectures that continue to be proposed recently~\cite{lu2017flexflow, gao2017tetris, mahmoud2018diffy, chen2018eyeriss}.

Regrettably, these proposals do not cover the complete space of dataflows at an exhaustive-enough level to serve as a reference for architects designing custom accelerators within a variety of constraints. In contrast, recent proposals on compilation~\cite{rstream:polyhedral,chen:tvm} and analysis tools~\cite{parashar2019timeloop} for DNNs analyze a broad space of software mappings of a DNN workload onto a given architecture, but the relationship between software mappings and hardware dataflows is not elucidated, and these black-box tools do not provide architects with intellectual intuitions on the consequences of dataflow selection and their impact on reuse. In fact, the very term "dataflow" is used in an inconsistent manner across both architecture and analysis proposals. Architects are thus left with an incomplete and unstructured set of intuitions on dataflows and the complex interplay between dataflow and microarchitecture choices.
In this paper, we seek to remedy this situation by providing a thorough set of insights on the choices and consequences of dataflow selection and their interplay with microarchitectural alternatives, and a structured mechanism to reason about them quantitatively. To that end, we make the following specific contributions.

First, we introduce a {\em data-centric} notation to represent various accelerator dataflows with data mappings and reuses being first-class entities, unlike the compute-centric notation used by prior proposals which infer the data reuses from a loop-nest representation~\cite{ma2017optimizing, chen2018eyeriss, lu2017flexflow, parashar2019timeloop}.
%
%
These data-centric directives can express a wide range of data-reuses (across space, time, and space-time) over arbitrary hierarchies of PEs for both dense and sparse DNN layers such as convolutions, LSTMs, and fully-connected layers.
%
%
We believe that our data-centric notation can complement the commonly used loop-nest notation, i.e., our notation can be viewed as an intermediate representation (IR) which can be extracted from a high-level loop-nest notation or specified directly.

Second, we show how these data-centric directives can be used to reason about reuse in a structured manner. We demonstrate the relationship between each directive, the specific form of algorithmic reuse exposed by the directive, and the potential ways to exploit that reuse using a hardware capability to improve efficiency. This analysis covers the complete space of ways in which any dataflow can exploit reuse.

Third, we introduce an analytical cost model named \tool (Modeling Accelerator Efficiency via Spatio-Temporal Reuse and Occupancy) that programmatically implements the above analysis. \tool takes as input 1) a DNN model with a set of layers, 2) a dataflow description for each layer specified using our proposed directives, and 3) the hardware configuration.  Based on these inputs, \tool outputs estimates of end-to-end execution time, energy (including all compute, buffer, and interconnect activities), NoC costs, and so on. 
A key challenge in our proposed approach is to provide a cost estimation that is both efficient and sufficiently precise to effectively support design space exploration.  We demonstrate \tool's abstract hardware model and analytic model to be within 90-95\% accuracy of actual open-source RTL~\cite{kwon2018maeri} while being 1029-4116$\times$ faster (10ms to run \tool versus 7.2-28.8 hours for an equivalent RTL simulation on a workstation with Xeon E5-2699 processor and 64GB memory).


Finally, we demonstrate how the \tool cost model can be used by accelerator designers to determine Pareto-optimal parameters for an accelerator with a given area, energy, or throughput budget. 
For a NVDLA~\cite{nvdla}-like dataflow (KC-Partitioned in~\autoref{table:EvalDataflows}) in VGG16~\cite{VGGnet} CONV layer 11, we see up to a 2.16$\times$ difference in power consumption between energy- versus throughput-optimized design points.
The energy-optimized design employs 10.6$\times$ more SRAM and 80\% the PEs of the throughput-optimized design.
This leads to an energy-delay product improvement of 65\%, with 62\% throughput.
The range of these numbers is a concrete example of the significance of this problem for accelerator architects.
\section{Background}
\label{sec:background}

\insertFigure{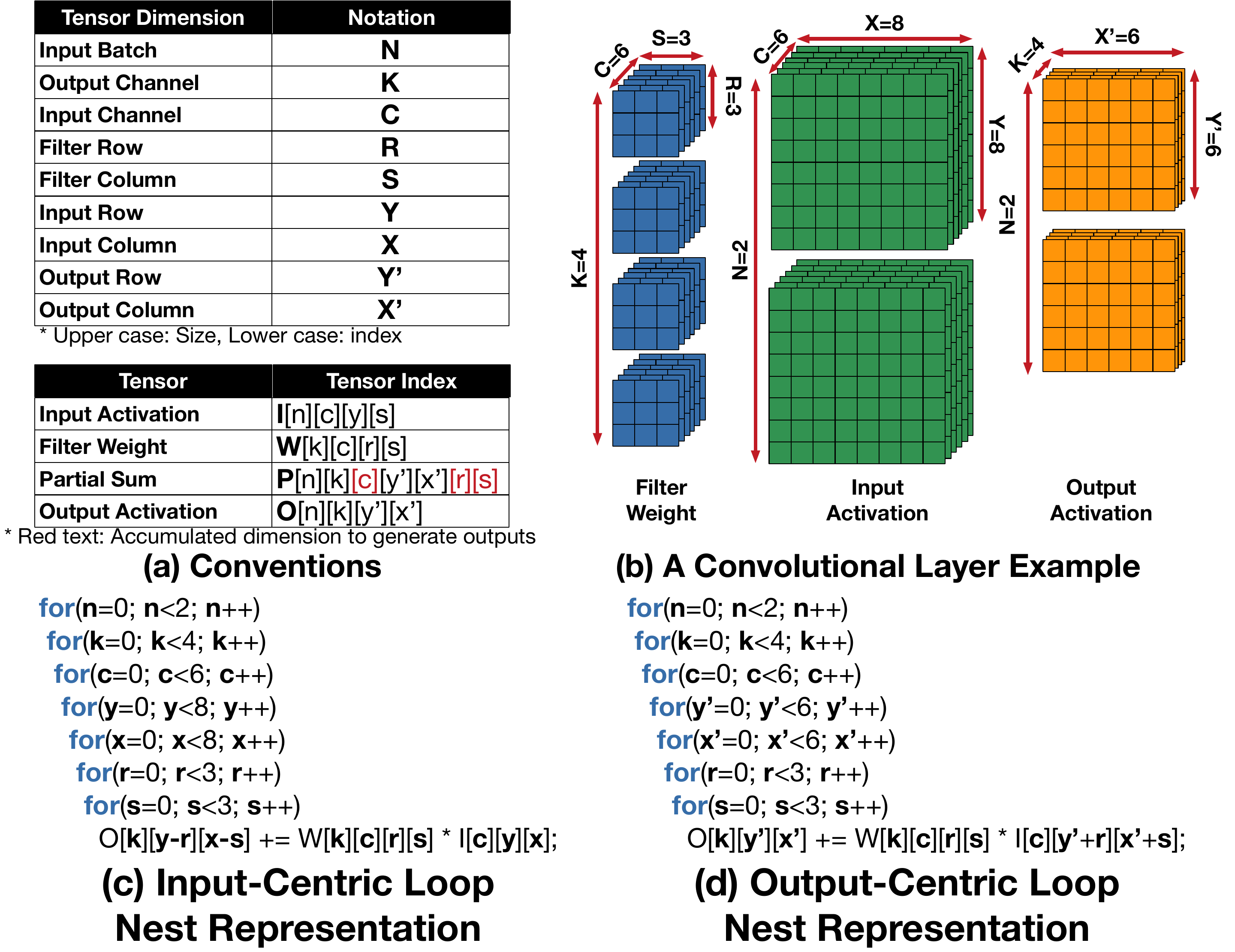}{Convolutional layer example}

To understand the cost-benefit tradeoffs of various approaches to compute convolutions, we discuss core concepts related to data reuse and dataflows in the context of DNN accelerators.

\subsection{Tensors in DNNs}
\label{subsec:tensors}

We present an example of a multi-channel 2D convolution in~\autoref{fig:7DConv_New} that involves seven data dimensions across three data structures: input/output activation and weight tensors.
Although our approach can be applied to various DNN layers---CONV2D, fully-connected (FC), LSTM, separable convolution, and so on---we focus on CONV2D and its variants in this paper because convolutional neural networks (CNNs) are popular, and CONV2D accounts for more than 90\% of overall computation in CNNs~\cite{cong2014minimizing, eyeriss_isca}.

Tensors in DNNs are addressed using seven dimensions in a complex manner.
For example, the row/column indices of output can be deduced using input row/column and filter row/column indices (i.e., an input-centric view of the convolution loop nest).
Also, the input channel index \texttt{c} appears in both filter and input activation, and the output channel \texttt{k} appears in both filter and output activation.
We call these dimensions {\it coupled} to these indices, as the position in the data space changes when the index is modified.
Because of these specific data access patterns, we can transform the loop nest to keep one of the data structures \emph{stationary} over a range of space or time (i.e., unchanged in a local buffer), which can significantly reduce global/local buffer access counts in DNN accelerators, as well as energy consumption by keeping local wires unchanging. 

\insertFigure{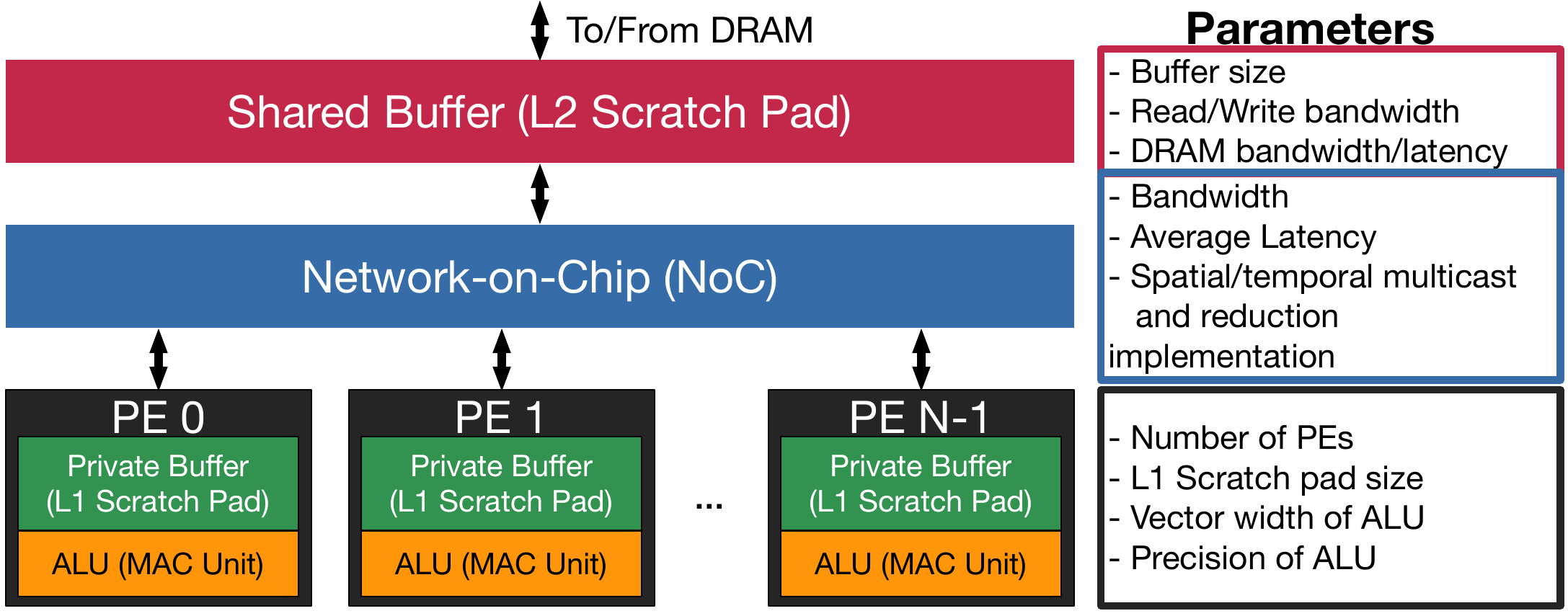}{Abstract DNN accelerator architecture model which is pervasive in many state-of-the-art accelerators~\cite{eyeriss_isca, sharma2016high, parashar2017scnn, jouppi2017datacenter, aklaghi2018snapea}. The illustrated base architecture can be hierarchically organized.  
}

\subsection{DNN Accelerators}
\label{subsec:DNNaccelerators}
DNN accelerators are specialized architectures to run DNN applications with high throughput and energy efficiency. As described in~\autoref{fig:HWModel}, most DNN accelerators employ hundreds of processing elements (PEs) to exploit inherent parallelism in DNN applications. PEs typically include scratchpad memories (L1) and ALUs that perform multiply-accumulate operations (MACs). To reduce energy- and time-consuming DRAM accesses, most DNN accelerators also include a shared scratchpad buffer (L2) large enough to stage data to feed all the PEs. Shared L2 buffer and PEs are interconnected with a network-on-chip (NoC). Our approach supports a wide range of interconnect designs in the NoC module. For example, a systolic array could be represented as a 2D array that provides unidirectional links toward East and South. 
Depending on the hardware parameters selected, our approach can support architecture designs that can efficiently execute a wide range of DNN operations, including convolutions, because it enables exploiting not only parallelism but also data reuse via buffers and forwarding/multicasting NoCs.

\subsection{Data Reuse Taxonomy}
\label{subsec:datareuse}

We observe that data reuse originates from two behaviors of DNN accelerators over time and space - multicasting (input tensors) and reduction (output tensors).

\betterparagraph{Multicasting} Spatial multicasting reads a data point from a buffer only once, spatially replicates the data point via wires, and delivers the data point to multiple spatial destinations (i.e., PEs), which reduces expensive remote buffer accesses and saves energy.
Likewise, temporal multicasting also reads a data point from a large remote buffer only once, temporally replicates the data point via a smaller local buffer, and delivers the data point to multiple temporal destinations (i.e., different time instances) at the same PE, which also reduces expensive remote buffer accesses and saves energy.

\betterparagraph{Reduction} Spatial reduction accumulates partial outputs from multiple spatial sources and spatially accumulates them via multiple compute units (e.g., an adder tree or reduce-and-forward).
Similarly, temporal reduction accumulates partial outputs from multiple temporal sources (i.e., partial sums computed at different time) and temporally accumulates them via an accumulation register or buffer (e.g., accumulation buffer in TPU~\cite{jouppi2017datacenter}).

\subsection{Dataflow Definition and Example}
\label{subsec:dataflowdefinition}

\insertFigure{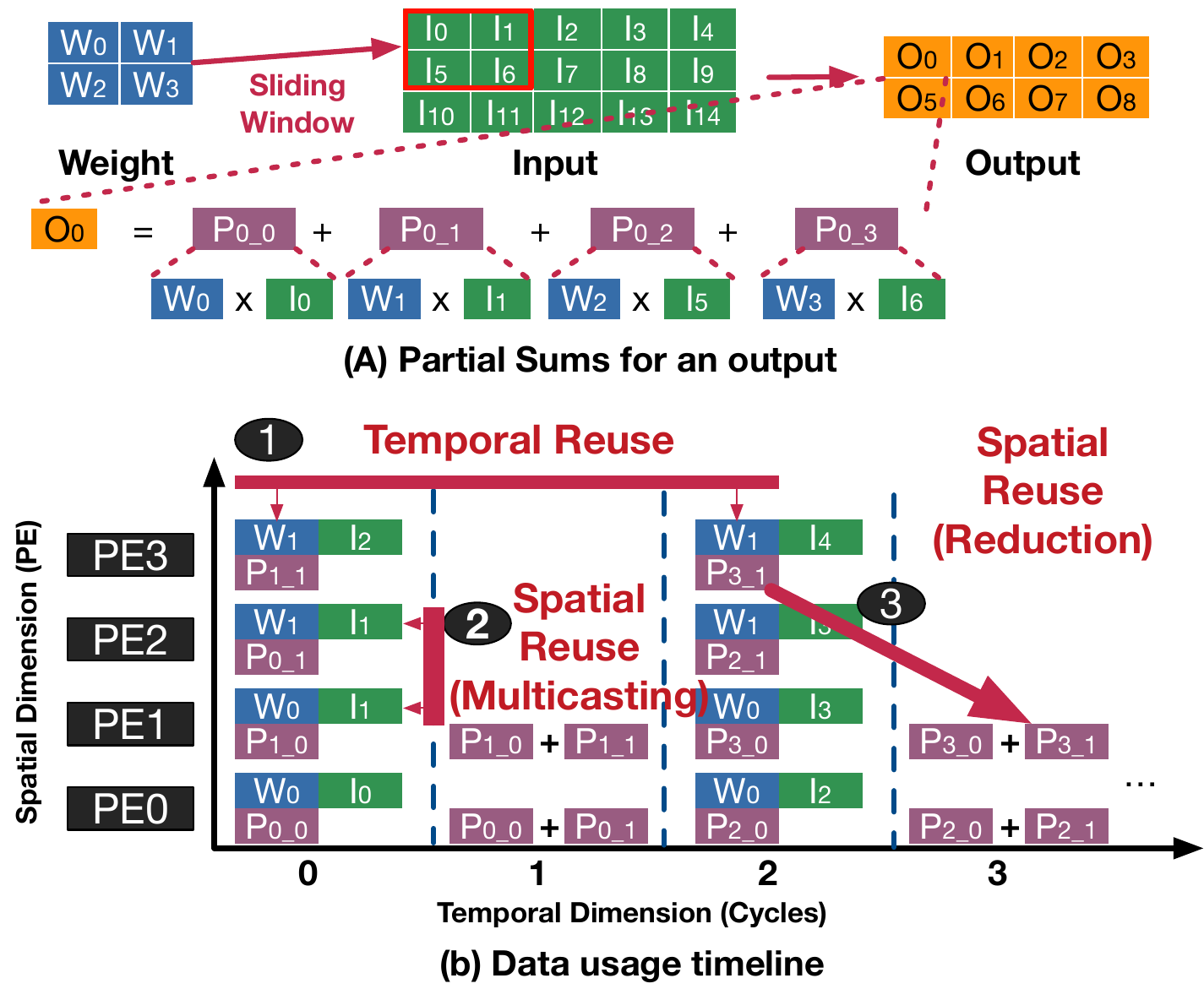}{An operational example of a weight-stationary style accelerator with four PEs. For simplicity, input/output channels and batch are omitted. A 2x2 kernel (R=2, S=2) is used in this example.}

In order to leverage these opportunities, the accelerator must schedule operations such that the PEs proceed through the data tensors in a coordinated fashion, which can be viewed as transformations (e.g., ordering and tiling) applied to the convolution in    \autoref{fig:7DConv_New}, along with a partitioning of data to PEs. 
Such schedules are termed as {\em dataflows} in prior work~\cite{eyeriss_isca}, which categorizes dataflows into classes based on the tensor which is scheduled to change least frequently, e.g., weight-stationary, output-stationary, and input-stationary.
%

\autoref{fig:DataReuse} shows an example weight-stationary dataflow run on four PEs.
We can observe that $W_{1}$ is multicast across time (temporal multicasting), $I_{1}$ is multicast across PEs (spatial multicasting), and $P_{3\_1}$ is reduced across space and time.
That is, the example accelerator temporally reuses $W_{1}$ and spatially reuses $I_{1}$ and $P_{3\_1}$.
Note that the name ``weight-stationary" conveys intuition and a high-level characterization of scheduling strategy, but detailed insight and analysis requires more precise description.

Chen et al.~\cite{chen2018eyeriss} refine the definition of dataflow by additionally specifying that two schedules which differ only in the concrete bounds should be considered \emph{instances} or \emph{mappings} of the same dataflow. 
This is an important distinction, as it allows families of accelerators to be categorized together even if they have different buffer sizes---i.e., a mobile chip and a datacenter chip may use the same traversal orderings despite large differences in tile size.
For brevity, for the remainder of this work, we make no distinction between schedules with fully-specified or partially unspecified concrete bounds but refer to them all as dataflows.

\subsection{Existing Expressions of Dataflow}
\label{subsec:dataflowdescription}

\insertWideFigure{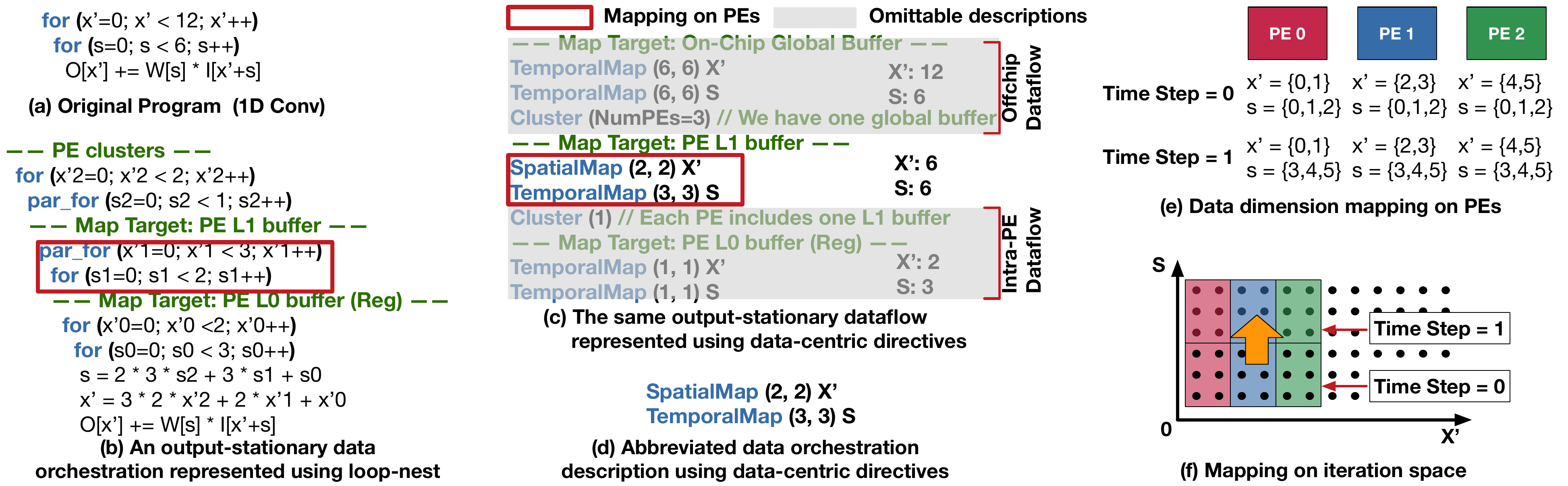}{An example 1D convolution and an example output-stationary dataflow on the convolution. We represent the dataflow (b) in loop nest and (c) data-centric directives. In (c), gray boxes represent omittable descriptions, which can be inferred (upper gray box) or do not affect the data reuse over PEs (lower gray box).  (d) shows an abbreviated form of the dataflow description in data-centric directives. (e) and (f) show resulting mapping on PEs and iteration space, whose dots correspond to computation (or, partial sums).}

To convey the scheduling decisions of a particular architecture, dataflows have been expressed as \emph{loop nests}, a syntax that resembles a simple imperative programming language with explicit parallelism, as presented in Eyeriss v2~\cite{chen2018eyeriss}. 
We term the loop nest notation a \emph{compute-centric} representation since the data movement is implicit from the loop order and the explicit parallelism specified by the programmer.
The loop order dictates the schedule (or, ordering) of computations, the explicit annotation of loops with {\tt parallel-for} captures parallelism, and the combination of loop ordering, tiling, and parallelism enables data reuse. 
Therefore, architects started to explore optimized loop nests encompassing all of the three aspects; loop order, parallelism, and tiling. For example, Eyeriss v2~\cite{chen2018eyeriss} describes its dataflow in a 22-dimensional loop nest.

Compute-centric representation including the polyhedral model has been a huge help to compilers in estimating reuses in guiding optimal loop transformations for both parallelism and locality~\cite{DBLP:journals/ibmrd/Sarkar97,DBLP:conf/ispass/SarkarM00,DBLP:conf/cc/ShirakoSFPRSS12,Wolf:1991:DLO:113445.113449,Bondhugula:2008:PAP:1375581.1375595,DBLP:conf/sc/PouchetBBCRS10,Shirako:2014:OWM:2683593.2683626}.
Those works provide sufficiently accurate cost estimations to drive a series loop transformation in a compiler.
However, they do not precisely model data reuse, so therefore computing throughput and energy-efficiency with high accuracy is challenging for those works.
Bao et al.~\cite{Bao:2017:AMC:3177123.3158120} developed an analytical model to accurately estimate cache behavior (thereby computing reuses) for a class of affine programs that can be precisely analyzed by a polyhedral model at compile time. 
However, they use heavyweight linear-algebra frameworks within the polyhedral model to compute reuse, thereby making it impractical to use these techniques on real large applications.
Also, it is very challenging for the polyhedral-based frameworks to compute reuse arising from array subscripts involving non-affine expressions or complex subscripts, such as modulus operations which are common in strided convolutions.

In addition, although there exists a body of past compiler work that performs reuse analysis on sequential programs~\cite{DBLP:journals/ibmrd/Sarkar97,DBLP:conf/ispass/SarkarM00,DBLP:conf/cc/ShirakoSFPRSS12,Wolf:1991:DLO:113445.113449,Bondhugula:2008:PAP:1375581.1375595,DBLP:conf/sc/PouchetBBCRS10,Shirako:2014:OWM:2683593.2683626,Bao:2017:AMC:3177123.3158120}, they lack the ability to analyze loop nests with explicit parallelism, while DNN dataflows often contain multiple levels of parallelism.
Also, those past works did not consider spatial reuse (which does not refer to the spatial locality in cache-based architectures but data reuse via wires or across PEs) that leverages multicasting and reduction support of accelerators, which plays a key role in estimating the overall throughput and energy efficiency of spatial DNN accelerators.

Such limitations and challenges motivate us to explore an alternative intermediate representation (IR) of dataflows, a \emph{data-centric} representation where data movement and organization are first-class entities.
Since data movement is explicit in the data-centric representation, our analytical model becomes simpler and relatively faster as there is no need to leverage heavyweight linear-algebra frameworks to precisely estimate data movement/reuse behavior.

\section{Describing Dataflows}
\label{sec:dataflow-IR}

\insertWideFigure{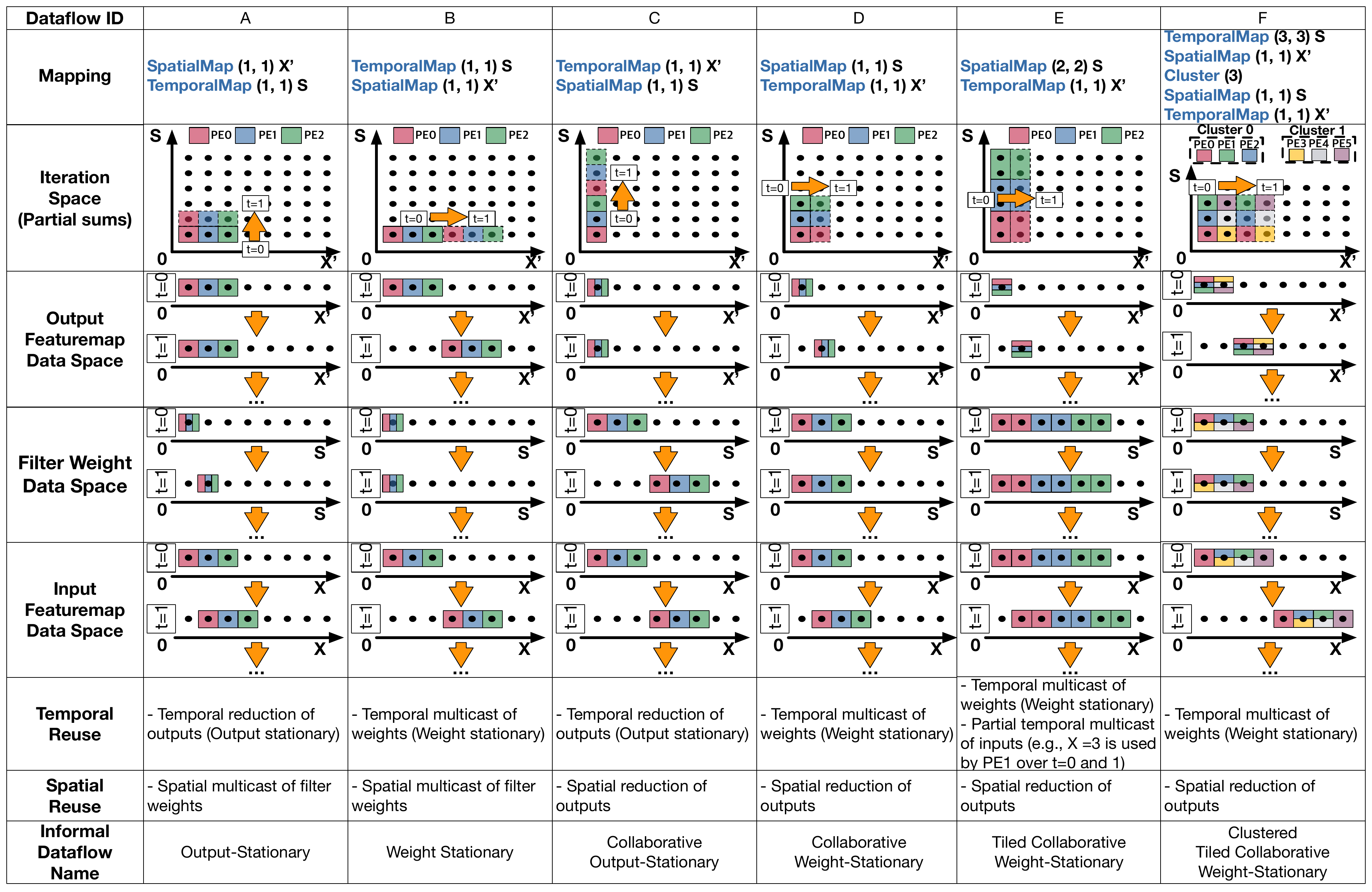}{The impact of directive order, spatial/temporal maps, tile sizes, and clustering over 1D convolution presented in~\autoref{fig:Unrolled1DConvExample}. The first row shows mapping described using our data-centric directives. The second row shows iteration spaces whose points correspond to each partial sum. In row three to five, we show data mapping of each data structure. Finally, we describe temporal and spatial reuse opportunities from each mapping.}

Our data-centric representation consists of four key directives -- 1) spatial map, 2) temporal map, 3) data movement order, and 4) clusters. 
We briefly explain all the key directives using 1D convolution (shown in~\autoref{fig:Unrolled1DConvExample} (a)) as a pedagogical example, and then discuss various hardware implementation choices for supporting a wide range of data-reuse across space, time, and space-time.

\subsection{Data-Centric Representation}
\label{subsec:dataflow-directives}

We define the dataflow of an accelerator design to consist of two major aspects -- (1) the schedule of DNN computations (e.g., choice of loop transformations) across time for exploiting a wide range of reuse, and (2) the mapping of the DNN computations across PEs for parallelism.
%
%
The representation is based on four key components, and we briefly discuss the first three components below. The fourth component, {\tt Cluster}, will be introduced in \autoref{subsec:dataflow_playground}.
\begin{enumerate}
    \item {\bf Spatial Map(size, offset) $\pmb{\alpha}$} specifies a distribution of dimension $\alpha$ (e.g., {\tt R}, {\tt X}) of a data structure across PEs, where {\tt size} refers to the number of indices mapped in the dimension $\alpha$ to each PE, and {\tt offset} describes the shift in the starting indices of $\alpha$ across consecutive PEs. 
    \item {\bf Temporal Map(size, offset) $\pmb{\alpha}$} specifies a distribution of dimension $\alpha$ of a data structure across time steps in a PE, and also the mapped chunk of dimension indices is the same across PEs in a time step. 
    The {\tt size} refers to the number of indices mapped in the dimension $\alpha$ to each PE, and {\tt offset} describes the shift in the starting indices of $\alpha$ across consecutive time steps in a PE. 
    \item {\bf Data Movement Order: } The sequence of spatial and temporal maps in the dataflow specification dictate the order of data movement, i.e., the change of the data mappings to PEs across time. 
\end{enumerate}

We demonstrate reuse opportunities presented by various dataflows using the 1D convolution example in ~\autoref{fig:Unrolled1DConvExample}(a).
We start by creating a unique dataflow for this program by the loop nest representation in 
~\autoref{fig:Unrolled1DConvExample}(b), assuming the accelerator has 2-level hierarchy (L0 register at PE + L1 local scratchpad buffer). 
The two loops enclosed in the red box are indicative of the mapping over the PEs, and their corresponding data-centric representation is in ~\autoref{fig:Unrolled1DConvExample}(c) and (d).

As can be seen from ~\autoref{fig:Unrolled1DConvExample}(e), the data elements corresponding to outputs (dimension {\tt X'}) is spatially distributed across three PEs, i.e., each PE receives different chunks of two output elements.
This particular data distribution can be captured with our spatial map directive with size and offset parameters being 2, resulting in {\tt SpatialMap(2,2) X'} where {\tt X'} is the first dimension of output data structure.
Also, the data elements corresponding to weights (dimension {\tt S}) is replicated across multiple PEs, i.e., each PE receives a same chunk of three weight elements in the first iteration, and receives different chunk of weight elements in the next iterations.
This particular replicated and temporal distribution can be captured with our temporal map directive with size and offset parameter being 3, resulting in {\tt TemporalMap(3,3) S}, where {\tt S} is the first dimension of the weight data structure.
Putting it together, spatial map on {\tt X'} followed by a temporal map on {\tt S} captures data mapping and movement behavior across PEs and time corresponding to the two loops in the loop-nest version, and these two directives are enclosed in the red box in~\autoref{fig:Unrolled1DConvExample}(c).
Each data-centric representation is a complete description of a unique dataflow.

\subsection{Dataflow Playground}
\label{subsec:dataflow_playground}

We build six example dataflows upon the simple 1D convolution discussed in~\autoref{fig:Unrolled1DConvExample} (d) to demonstrate how small changes to a dataflow expose various forms of reuse---both spatial and temporal.
~\autoref{fig:Dataflow_Examples} illustrates those six example dataflows, which consists of a base dataflow~\autoref{fig:Dataflow_Examples}(A) and its variants.
We modify the directive order, spatially/temporally mapped dimensions, mapping size, and PE clustering and discuss their impact on data reuse.

\betterparagraph{Directive Order}
A change in directive order can result in an entirely different temporal reuse (or, stationary behavior).
For example, the sequence of directives in mapping in~\autoref{fig:Dataflow_Examples}(A) indicates that all data indices of S should be explored before working on the next chunk of {\tt X'} indices. 
This order results in temporally reusing values of data corresponding to  {\tt X'}  indices (i.e., partial sums) for all indices of {\tt S}.
Therefore, this dataflow is informally referred to as output-stationary and partitioned across multiple outputs in parallel.  
%
%
\autoref{fig:Dataflow_Examples}(B) shows the impact of interchanging the order of directives. This results in a weight-stationary dataflow, because PEs can temporally reuse weight values corresponding to {\tt S} indices, for all indices of {\tt X'} before going to next chunk of {\tt S} indices.
Similarly, \autoref{fig:Dataflow_Examples}(C) and (D) shows the spatial distribution on {\tt S} instead of {\tt X'}, and also the impact of data movement order on temporal reuse leading to different dataflow variations. This indicates why the informal dataflow name should not be taken as a complete and precise specification of its behavior.

\betterparagraph{Spatially and Temporally Mapped Dimensions} 
%
%
In~\autoref{fig:Dataflow_Examples}(A) the directive {\tt SpatialMap(1,1) X'} (where {\tt X'} refers to the first dimension of the output data structure), spatially distributes indices of the {\tt X'} dimension with a chunk size of one (the {\tt size} parameter) across PEs with an offset of one (the {\tt offset} parameter).
This means that each PE works on a different column of the output data space.
If the number of PEs is not sufficient to cover all indices of the dimension mapped, then the mapping is folded over time across the same set of PEs. 
Also, if {\tt offset} value is smaller than {\tt size} value, then there will be an overlap of indices across consecutive PEs, and this is useful in describing mappings on input activation dimensions X and Y because their iteration space is skewed.

%
Similarly, {\tt TemporalMap(1,1) S} (where {\tt S} refers to the first dimension of filter weight data structure), distributes indices of the {\tt S} dimension with a chunk size of one across time steps with an offset of one.
This means that each PE works on the same column of the weight data space.
Since all PEs get the same data indices corresponding to a temporally mapped dimension, this creates an opportunity for {\em spatial reuse}, i.e., multicasting the same data values across PEs in a time step.

\betterparagraph{Mapping Size}
In all of the mappings from \autoref{fig:Dataflow_Examples}A-D, the mapping sizes (first argument) of weights and outputs are one -- resulting in full temporal reuse of weights but no temporal reuse of outputs (e.g., mapping B and D) or vice versa (e.g., mapping A and C). There is no temporal reuse of inputs in any mapping.
%
Increasing the map size of the spatial or temporal maps can help in presenting opportunities for partial temporal reuse, which can capture convolutional reuse of inputs in CNN layers. 
For example, the spatial map corresponding to the {\tt S} dimension in \autoref{fig:Dataflow_Examples}(E) helps in exploiting the partial temporal reuse of input data across time steps.

\betterparagraph{PE Clustering for Multi-dimensional Spatial Distributions}
As can be seen in \autoref{fig:Dataflow_Examples}(A-E), data mappings related to a map in the outer position get updated after a full exploration of a map in the inner position. 
%
%
This inherent assumption can limit certain dataflow behaviors where one might be interested in simultaneously exploiting spatial distribution of more than one data dimensions.

To address this, we introduce another directive called {\em Cluster} as a mean to support the simultaneous spatial distribution of multiple data dimensions.
The cluster directive logically groups multiple PEs or nested sub-clusters (when a dataflow has multiple cluster directives) of {\tt size} parameter.
For example, \Cluster(3) in ~\autoref{fig:Dataflow_Examples}(F) arranges available PEs into groups of three, resulting in two clusters of three PEs.

All the mapping directives specified above a \Cluster directive perform the mapping across logical clusters created by the \Cluster directive.
All the mapping directives specified below a \Cluster directive perform the mapping across PEs or lower level logical clusters inside a logical cluster created by the \Cluster directive.
That is, all the mapping directives above a \Cluster directive see logical clusters while those below the \Cluster directive see \textit{inside} of each logical cluster.
%
%
With this mechanism, one can specify complex dataflows with multiple parallelization dimensions represented by multiple \sMap directives (one in each cluster level).
An example of this can be seen in~\autoref{fig:Dataflow_Examples}(F), where the {\tt X'} dimension is spatially distributed across clusters, and the {\tt S} dimension is spatially distributed within the cluster. 
The cluster directives enable us to represent existing real-world accelerator dataflows, such as Eyeriss~\cite{eyeriss_isca} since it involves the spatial distribution of R and Y dimensions simultaneously, and also NVDLA~\cite{nvdla} which involves the spatial distribution of K and C dimensions. 
Another advantage of the cluster directive is that its notion of grouping multiple PEs can represent coarse-grained PEs in accelerators, such as SIMD units~\cite{song2016cbrain} and matrix tensor accelerators like GPU Tensor Cores.

In summary, we discussed five transformations 
that capture all possible aspects of dataflows: scheduling, tiling, and mapping.
As shown in~\autoref{fig:Dataflow_Examples} the data-centric directives can concisely represent all of those aspects.
%
%
We envision that the data-centric representation could be either auto-generated from a loop nest version of the dataflow (with affine constraints), or manually written.






\insertWideTableFigure{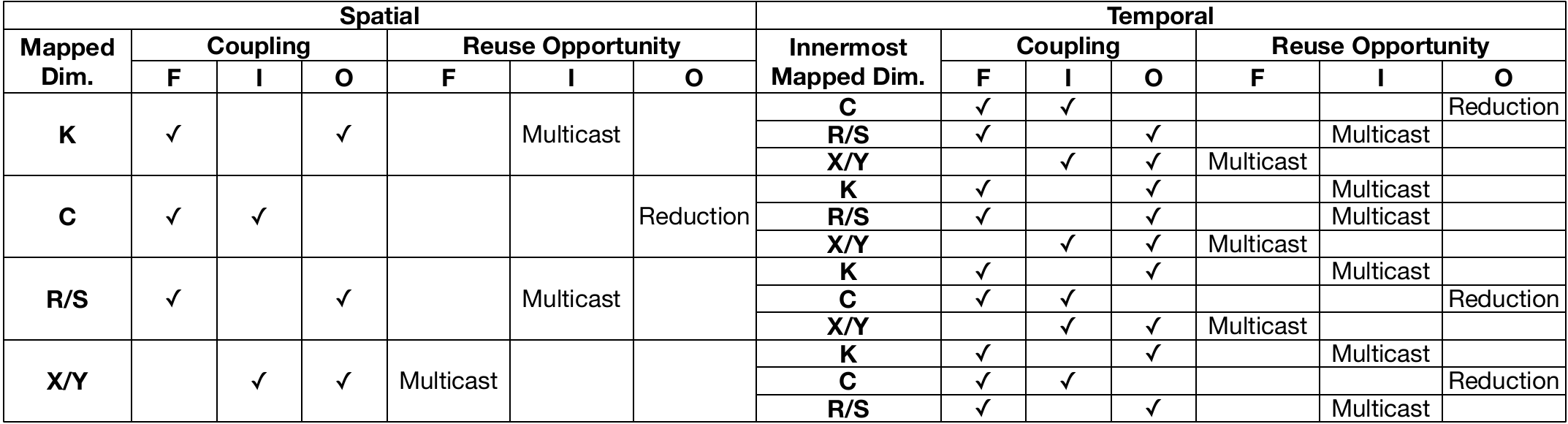}{Reuse opportunities based on spatially-mapped dimensions in combination with innermost temporally-mapped dimensions. Filters (F), Inputs (I), and Outputs (O) are considered separately. For brevity, X/Y should be interpreted as X'/Y' as appropriate.}

\insertTableFigure{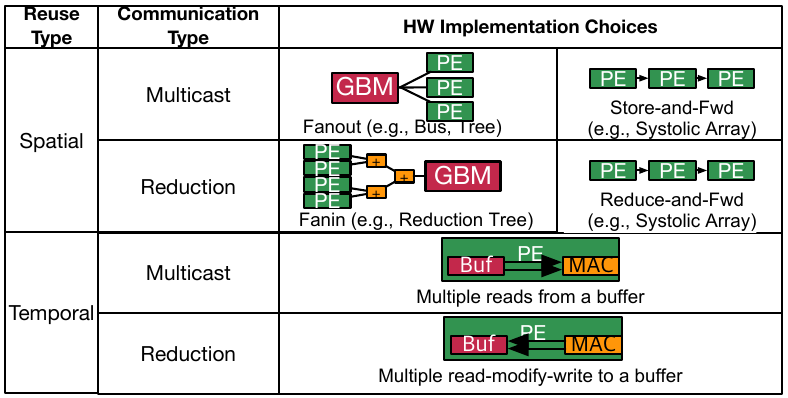}{Hardware Implementation Choices for supporting spatial and temporal reuse. Note - by {\it temporal multicast},  we refer to {\it stationary} buffers from which the same data is read over time.}

\subsection{Hardware Implications of Reuse}
\label{subsec:hardware_implementation}

As we discussed above, various data reuse opportunities appear based on the dataflow.
~\autoref{table:DataReuseOpportunities} summarizes how such opportunities appear in the relationship of spatially mapped dimension within a cluster (Map column) and inner-most temporally mapped dimension (InnerMap column).
%
%
For example, if output channels (K) are spatially mapped, a decoupled data structure, input feature map, does not change over space.
That is, all the PEs receive the same input feature map, which implies a full spatial reuse opportunity (broadcast).
In the same example, when the inner-most temporally mapped dimension is the input channels (C), the input channel changes every iteration, which provides temporal reduction opportunities of outputs.

Although a dataflow provides temporal or spatial data reuse opportunities, appropriate hardware support is required to actually exploit these phenomena.
~\autoref{table:HW_Impl_choices} summarizes four reuse categories and corresponding hardware implementation to support them.
As the table shows, reuse can be either spatial or temporal.
Based on the data structure, the communication type can be either multicast (input tensors) or reduction (output tensors).
Multicast is a communication type that delivers the same data to multiple targets over space (different PEs at the same time) or time (the same PE in different time).
Therefore, multicast is one to many communication type, which requires either a fan-out network-on-chip structure such as bus or tree, or a ``stationary" buffer to hold the data and deliver it to the future.
In contrast, the reduction is many to one communication type, which applies to partial sums to generate final outputs.
The reduction also can be either spatial or temporal.
Example hardware to support spatial reduction is a reduction tree or reduce-and-forward chain such as systolic arrays.
Temporal reduction can be supported by a read-modify-write buffer.

In summary, different dataflows (expressed via our directives) expose different forms of reuse: spatial and temporal, both for multicasts and reductions, which in turn can have multiple hardware implementations.
Reasoning about dataflows in this structured manner exposes new insights and potential microarchitectural solutions.
The discussion so far focused on a simple 1D convolution, which itself exposed many possible dataflows and reuse opportunities.
We extend this to a full convolution loop and analyze reuse opportunities within a specific dataflow.

\subsection{Extended Example: Row-stationary Dataflow}

\insertWideFigure{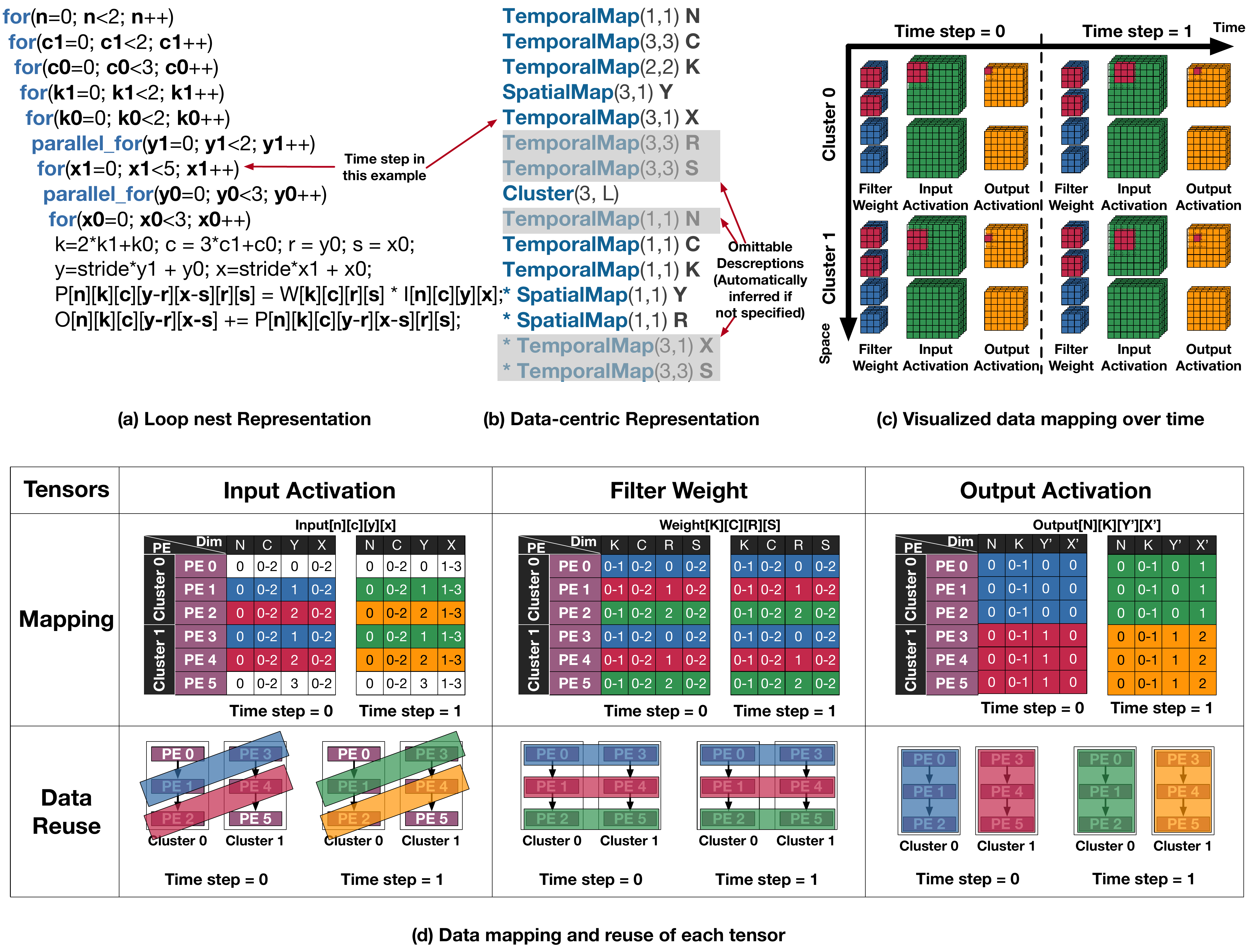}{An extended example of a row-stationary style dataflow mapped on a six-PE accelerator. We select our own tile sizes for any not specified in the original work~\cite{eyeriss_isca}. We do not apply additional mapping optimizations to minimize PE under-utilization. Colors represent data replication either across time or space (PEs). Directives with asterisks indicate fully unrolled directives that cover entire data dimension with one mapping.}

~\autoref{fig:EyerissDeepDive} presents detailed mapping and reuse patterns across two unit time steps of an example row-stationary dataflow~\cite{eyeriss_isca} over a six-PE accelerator.
The accelerator has two PE clusters with three PEs in each cluster.
We use the same example layer previously used in~\autoref{fig:7DConv_New}.
~\autoref{fig:EyerissDeepDive}(a) and (b) are compute- and data-centric representations of the row-stationary dataflow.
~\autoref{fig:EyerissDeepDive}(c) shows how the mapping moves across space (PE clusters) and time
~\autoref{fig:EyerissDeepDive}(d) shows the actual coordinates of each tensor across two time steps and two clusters (i.e., time and space).
Each colored box in~\autoref{fig:EyerissDeepDive}(d) represents replicated data points, which imply reuse opportunities.
Based on the replicated data points, we can infer data reuse over the PE array, as shown in data reuse row in~\autoref{fig:EyerissDeepDive}(d).
The mapping in~\autoref{fig:EyerissDeepDive}(d) shows that the same set of input activation values are replicated across two clusters in a skewed manner within the same time step, which implies spatial reuse opportunities in the diagonal direction of the example PE array.
Similarly,~\autoref{fig:EyerissDeepDive}(d) shows that the same set of weight values are replicated over two time steps within the same PE, which implies temporal reuse opportunities and weight-stationary style dataflow in unit time step granularity.
Note that the dataflow is still row-stationary in a coarse-grained time step although it is weight stationary in unit time steps we define in ~\autoref{fig:EyerissDeepDive} (a) and (b).
Finally,~\autoref{fig:EyerissDeepDive} (d) shows the same set of output activation over PEs in each PE cluster, which means that all the PEs in each cluster cooperate to generate a set of output activation data.
That is, each PE in a PE cluster generates different partial sums for the same output activation, and they need to be accumulated across PEs in each PE cluster to generate final output activation values.
Based on the example analysis in~\autoref{fig:EyerissDeepDive}, we observe that the data reuse pattern exactly matches the original work~\cite{eyeriss_isca}: reuse in the horizontal direction for filter weights and vertical for outputs (partial sum accumulation), and reuse in the diagonal direction for input activations.

In summary, reuse opportunities are based on the replicated data across time or space (PEs), which implies temporal and spatial reuse opportunities, respectively. The examples in this section demonstrate the need for a fast, accurate quantitative methodology to compute reuse for complex dataflows.
 
\section{Quantitative Dataflow Analysis}
\label{sec:framework}


\insertWideFigure{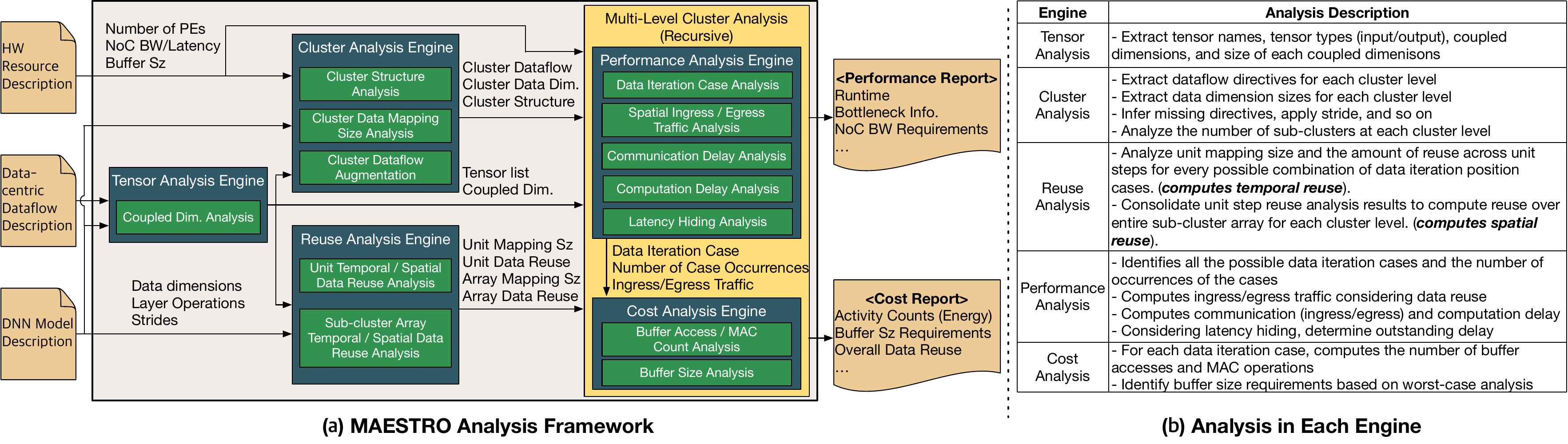}{An overview of \tool's analysis framework. For simplicity, we omit components other than analysis engines.}

\insertFigure{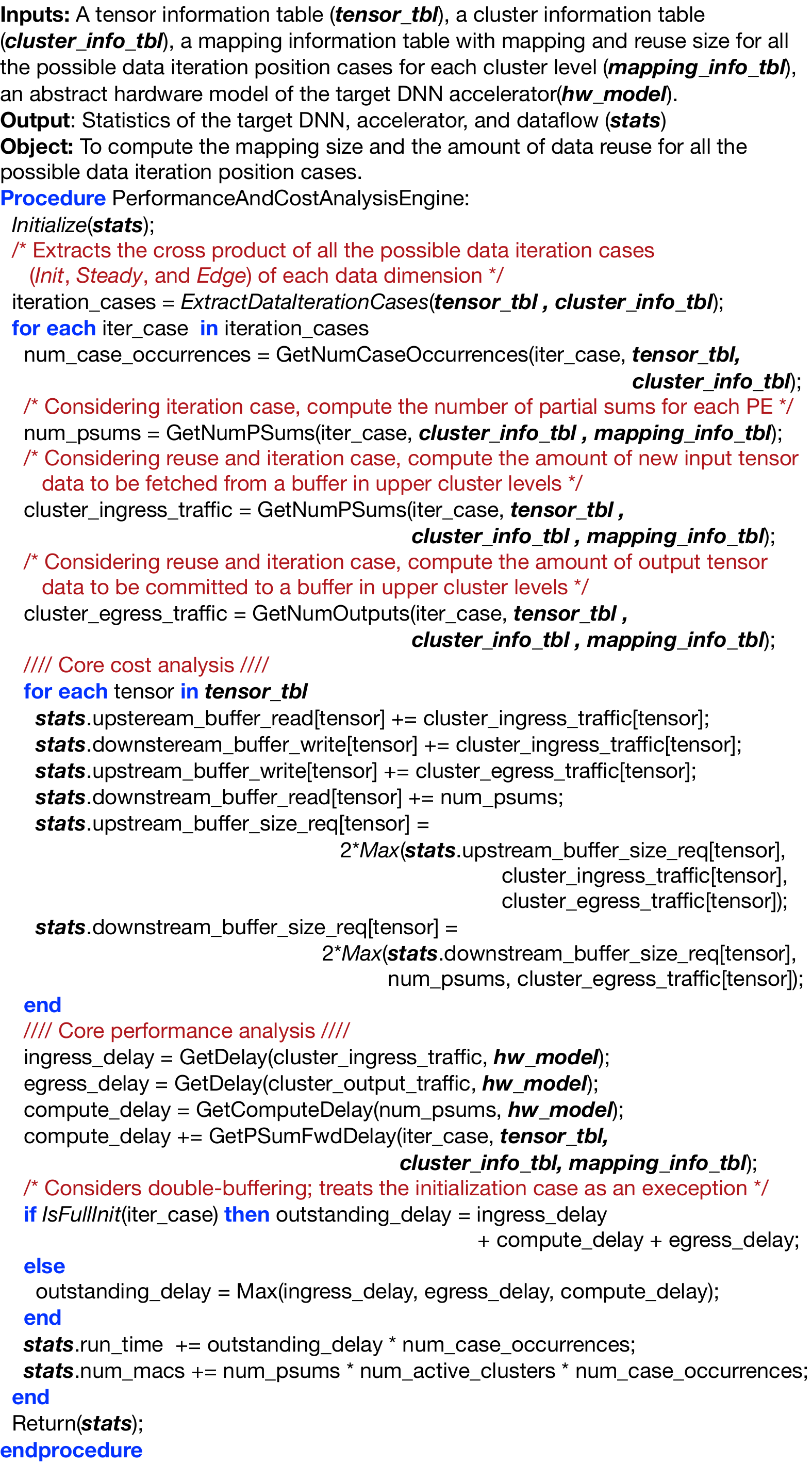}{A high-level overview of algorithms in performance and cost analysis engines.}

In this section, we present our approach to quantitatively estimating runtime and energy efficiency of dataflows on a target DNN model and hardware configuration.
Based on the approach, we implement an analysis framework, \tool, which consists of five engines: tensor, cluster, reuse, performance analysis, and cost analysis.
~\autoref{fig:AnalysisFramework} provides a high-level overview of the five engines.
In the interest of space, we only discuss high-level algorithms without edge case handling, multiple layers, and multiple cluster levels.
For details, we present them in our open-source repository~\cite{maestro_opensource}.

\subsection{Preliminary Engines}
\label{subsec:prelim_engines}

\betterparagraph{Tensor Analysis}
As described in~\autoref{fig:AnalysisFramework}, the tensor analysis engine identifies dimension coupling for each tensor based on specified layer operations.
For example, in depth-wise convolutions, output activation is not coupled with the output-channel dimension but coupled with the input channel dimension.
Note that depth-wise convolution can be understood either in this manner or by eliminating input channel dimension (C).
We select this convention because it aligns with \tool's input-centric cost model.
\tool allows users to specify tensors with arbitrary dimension coupling, and such coupling relationship is input to the rest of engines, which provides generality to \tool.

\betterparagraph{Cluster Analysis}
A PE cluster refers to a group of PEs that processes one or more data dimensions in parallel, specified by the \Cluster directive.
~\autoref{fig:AnalysisFramework} (b) describes the analysis in Cluster Analysis (CLA) engine.
The CLA engine analyzes a given dataflow description written in dataflow directives to identify the number of sub-clusters, extract cluster dataflow directives and data dimensions, and augment the given dataflow descriptions for missing directives, stride handling, and so on, for each cluster level.

\betterparagraph{Reuse Analysis}
~\autoref{fig:AnalysisFramework} (b) includes a high-level description of analysis in data reuse analysis (RA) engine.
RA engine identifies the amount of temporal and spatial reuse across adjacent time steps, which is the data iteration corresponding to the inner-most non-temporally/spatially unrolled mapping directive.

\subsection{Performance Analysis}
\label{subsec:performance-analysis}

~\autoref{fig:AnalysisFramework} (a) presents a high-level overview of the performance and cost analysis engine, and~\autoref{fig:PerformanceCostAnalysisEngine} shows high-level algorithm of the performance analysis (PA) engine.
Utilizing the reuse information computed in the RA engine, PA engine computes the runtime for all the possible cases based on the data dimension and dataflow.
The computed runtime is multiplied with the number of each case's occurrences and accumulated to compute the total runtime.
The runtime of a DNN accelerator consists of communication delay (L2 to L1, L1 to L2, local forwarding) and computation delay in each PE, which are directly related to the accelerator's hardware parameters.
PA engine considers double buffering when it computes the outstanding delay (the worst case delay of communication/computation delay) that directly contributes to the runtime.

To estimate communication delays, \tool relies on its analytical network-on-chip (NoC) model based on a pipe model similar to other analytic models~\cite{parashar2019timeloop}.
The pipe model utilizes two parameters, the pipe width (bandwidth) and length (average delay), to estimate the communication delay via NoC.
The model incorporates a pipelining effect as many packet-switching NoCs have similar behavior.
Various combinations of the bandwidth and average delay enables to model NoC structures with reasonable accuracy.
For example, Eyeriss~\cite{eyeriss_isca} has a two-level hierarchical bus with dedicated channels for input, weight, and output tensors.
Therefore, a bandwidth of 3X properly models the top level NoC.
The average latency depends on implementation details; users should choose an appropriate value considering implementation details (e.g., the use of ingress/egress buffers, which adds one cycle delay each).
For more complicated NoC architectures, users should select bisection bandwidth and average latency considering uniform communication to all the PEs from a global buffer.
For example, a $N \times N$ 2D mesh network with the injection point at one of the corners, the bisection bandwidth is $N$, and the average latency is $N$.
Assuming that the user has access to the NoC implementation information, the NoC model is precise when the NoC is a bus or a crossbar.

\subsection{Cost Analysis}
\label{subsec:cost-analysis}

~\autoref{fig:PerformanceCostAnalysisEngine} describes how the cost analysis (CA) engine computes the number of buffer accesses and estimates the buffer size requirements for each tensor, considering data reuse computed in the RA engine and data iteration cases.
Utilizing the access counts and the number of MAC operation information, \tool computes the energy cost.
\tool includes an energy model based on those activity counts and Cacti~\cite{muralimanohar2009cacti} simulation, which can be replaced by any other energy model based on such activity counts (e.g., Accelergy~\cite{iccad_2019_accelergy}).

\subsection{Complex Dataflow Analysis}
\label{subsec:complex_dataflow_analysis}

\betterparagraph{Multi-cluster Analysis}
Multi-cluster cases can be split into single-cluster cases with the data dimension size set as the mapping size of the corresponding mapping directive in the upper cluster.
The outstanding delay of a cluster level becomes the computation delay of the next cluster level above.
To handle various edge cases that affects all the lower cluster levels, \tool recursively performs performance and cost analysis, as illustrated in~\autoref{fig:AnalysisFramework}. 
In the recursive analysis, the base case is the inner-most cluster whose sub-clusters are actual PEs.
Although \tool performs recursion, the complexity is not high because the number of PE cluster levels are typically two or three.
%
%
Note that each of the edge cases at each cluster level also needs to be recursively processed.
However, in most cases, we observe the number of edge cases across cluster levels is less than 20, which is still in a tractable scale.

\betterparagraph{Other DNNs}
Although we used dense convolution as examples for simplicity, \tool can model a variety of layers (LSTM hidden layer, pooling, fully-connected, transposed convolution, and so on) based on the generality of the data-centric approach.
Our data-centric approach supports all the operations represented as the loop nest with two input tensors and one output tensor wherein all the tensor indices are coupled in only one or two data dimensions in affine functions.
\tool also can model uniformly distributed sparsity for any supported dataflow. 
Support for more complex statistical sparsity distributions is future work.

\subsection{Model Validation}
\label{subsec:validation}

\insertFigure{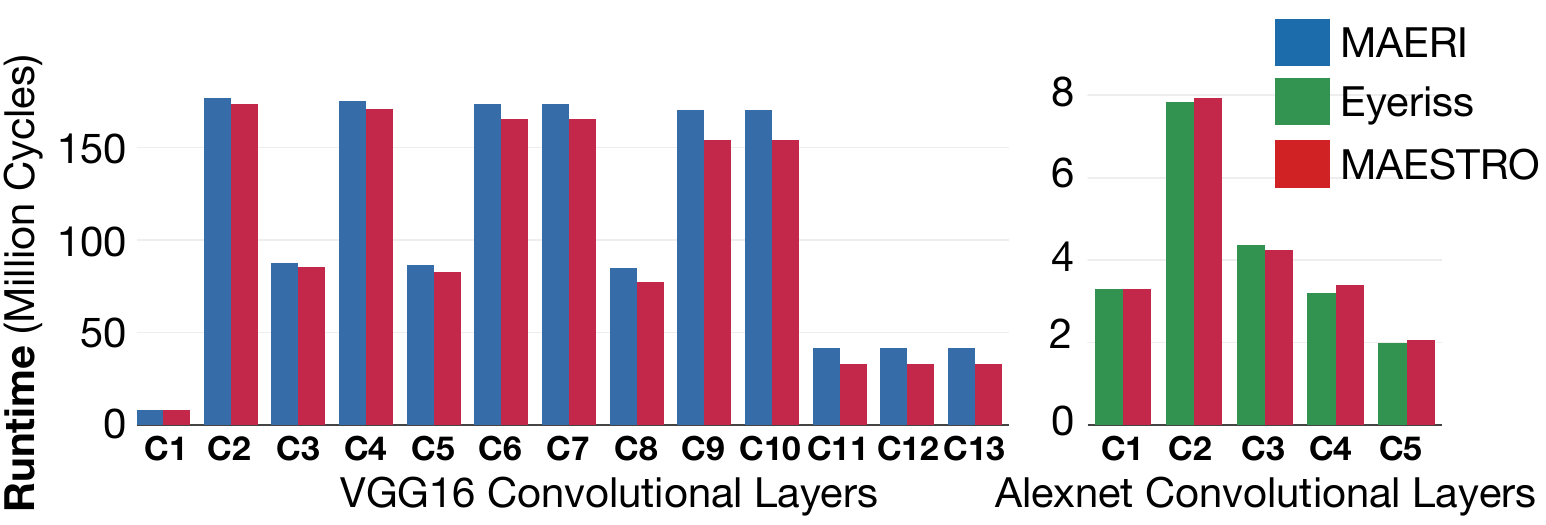}{Runtime model validation against MAERI~\cite{kwon2018maeri} RTL simulation with 64 PEs and Eyeriss~\cite{chen2017eyeriss_issc} runtime reported in the paper with 168 PEs.}

We validated \tool's performance model against RTL simulations of two accelerators - MAERI~\cite{kwon2018maeri} and Eyeriss~\cite{chen2017eyeriss_issc} when running VGG16 and AlexNet respectively\footnote{MAERI RTL is open-source. For Eyeriss, we use the reported runtime for AlexNet because detailed mapping parameters are described for only AlexNet in the paper.}.
\autoref{fig:Validation} shows that the runtime estimated by \tool are within 3.9\% absolute error of the cycle-accurate RTL simulation and reported processing delay~\cite{chen2017eyeriss_issc} in average.
\section{Case Studies}
\label{sec:eval}

\insertTableFigure{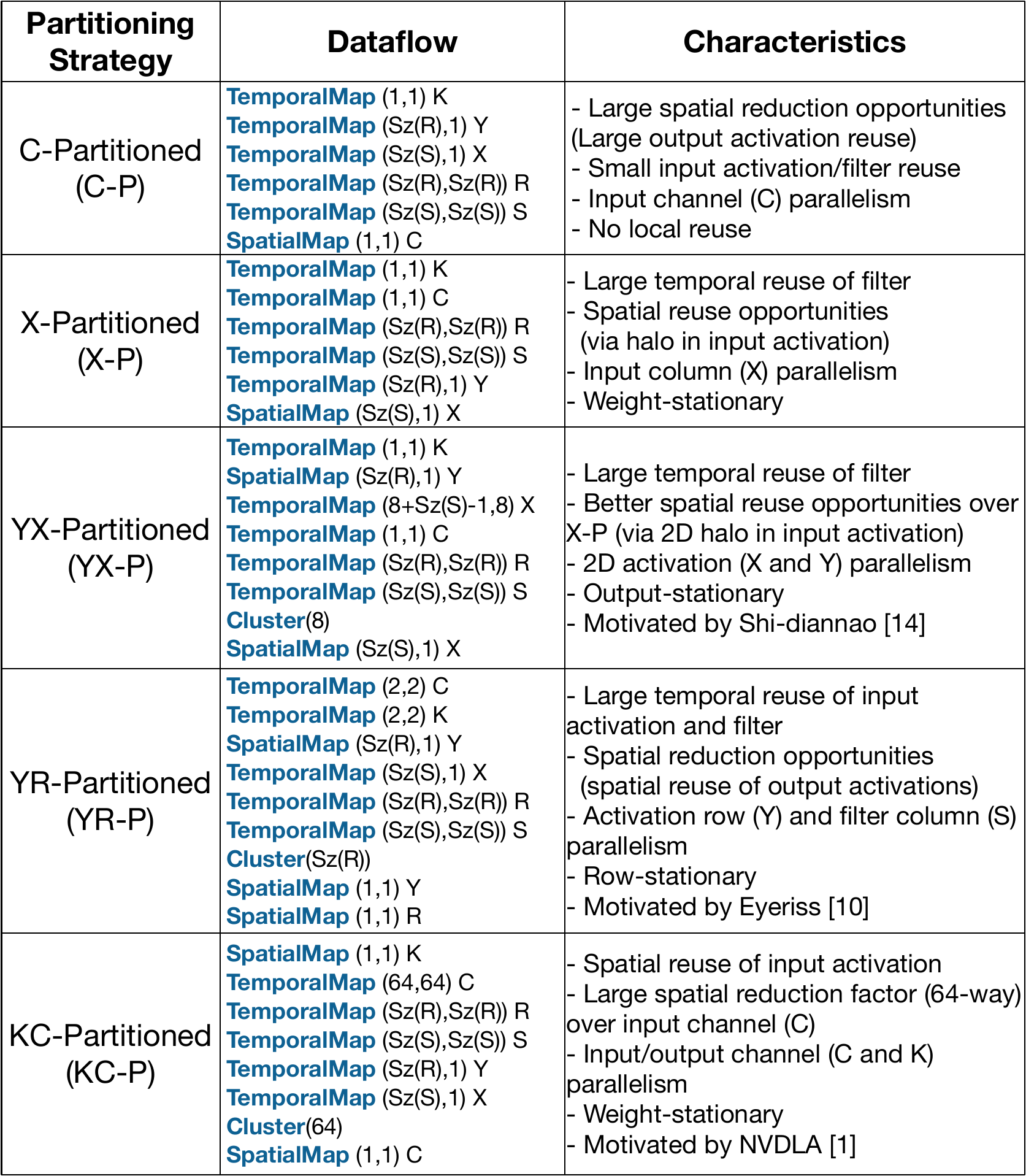}{Five example dataflows used for the evaluation. For conciseness, we omit redundant directives that are automatically inferred by \tool. YX-P, YR-P, and CK-P dataflows are motivated by Shidiannao~\cite{du2015shidiannao}, Eyeriss~\cite{eyeriss_isca}, and NVDLA~\cite{nvdla}, respectively. The name of each dataflow is based on spatial dimensions from the upper-most cluster level.
}

\insertWideFigure{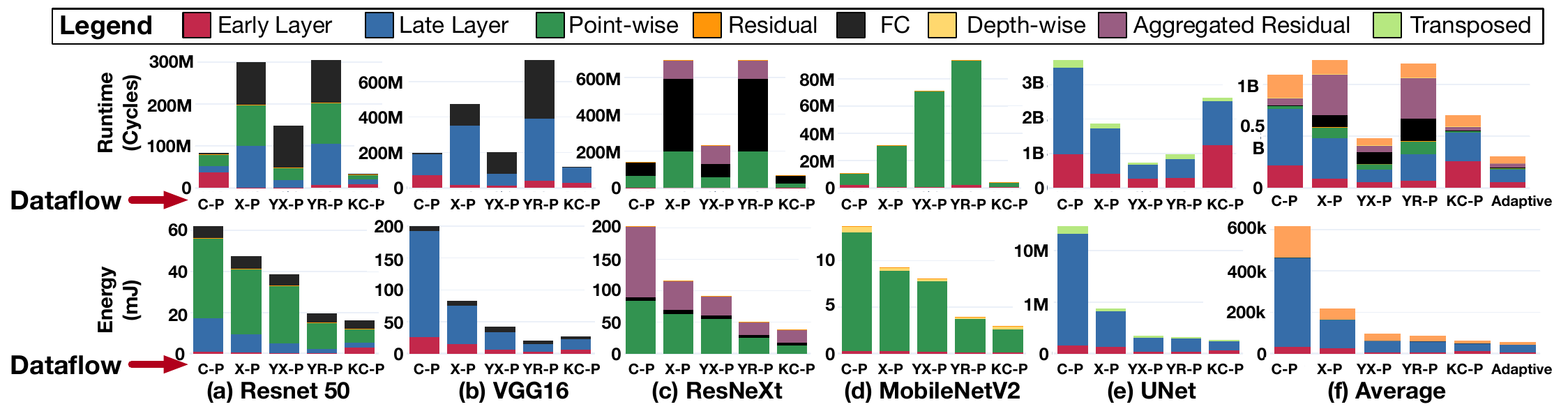}{Plots in top and bottom rows present runtime and energy estimation of five dataflows listed in the table, respectively. We apply 256 PEs and 32GBps NoC bandwidth. We evaluate all the dataflows using five different DNN model; Resnet50~\cite{Resnet}, VGG16~\cite{VGGnet}, ResNeXt50~\cite{ResNeXt}, MobileNetV2~\cite{mobilenetv2}, and UNet~\cite{UNet}. The final column (f) presents the average results across models for each DNN operator type listed in~\autoref{table:DNN_Operators} and the adaptive dataflow case.}

\autoref{table:DNN_Operators} summarizes the features of frequently used DNN operators from state-of-the-art DNN models~\cite{Resnet, ResNeXt, mobilenet, mobilenetv2, UNet}.
Early and late layers refer to layers with high-resolution activation with shallow channels and vice versa, respectively.
We label them as early and late layers because such layers appear early and late in classification networks~\cite{Resnet, ResNeXt, mobilenetv2, VGGnet}.
We compare the number of input channels and the input activation height to identify them\footnote{If C > Y, late layer. Else, early layer}.

With \tool, we perform deeper case studies about the costs and benefits of various dataflows when they are applied to different DNN operations listed in~\autoref{table:DNN_Operators}.
We evaluate five distinct dataflow styles listed in~\autoref{table:EvalDataflows} in~\autoref{subsec:dataflow_case_study} and the preference of each dataflow to different DNN operators.
For energy estimation, we multiply activity counts with base energy values from Cacti~\cite{muralimanohar2009cacti} simulation (28nm, 2KB L1 scratchpad, and 1MB shared L2 buffer).
We also present distinct design space of an early layer (wide and shallow) and a late layer (narrow and deep) to show the dramatically different hardware preference of different DNN operator styles and dataflow in~\autoref{subsec:dse_eval}.

\subsection{Case study I: Dataflow Trade-offs}
\label{subsec:dataflow_case_study}

\insertTableFigure{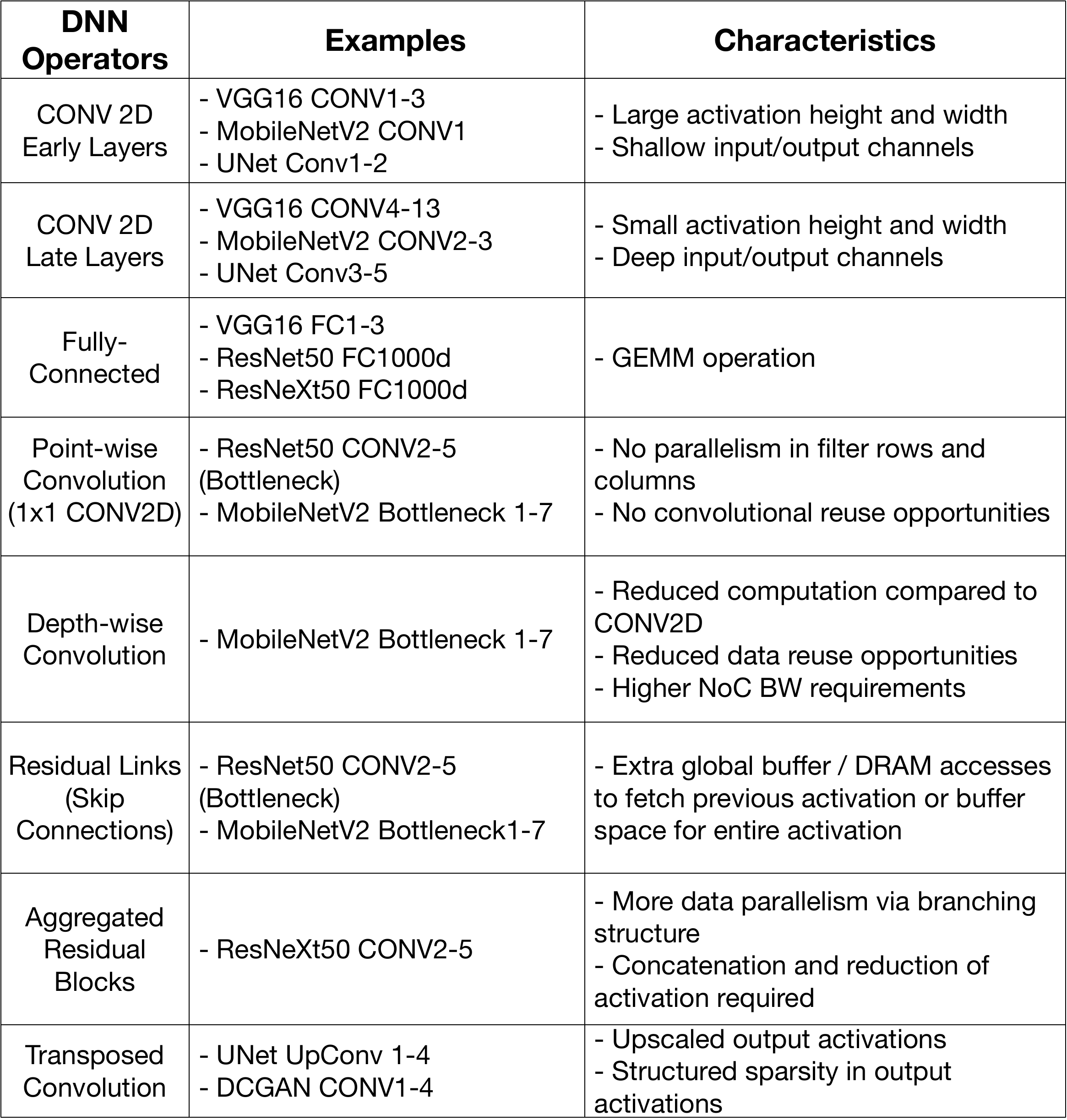}{Operators in state-of-the-art DNNs and their features and implication. Bottleneck~\cite{Resnet} and depth-wise separable convolution~\cite{mobilenet} are listetd in a fine-grained operators (point-wise convolution, depth-wise convolution, and residual links). Examples are based on notable networks (VGGnet~\cite{VGGnet} and DCGAN~\cite{DCGAN}) and state-of-the-art networks (MobileNetV2~\cite{mobilenetv2}, ResNet50~\cite{Resnet}, ResNeXt50~\cite{ResNeXt}.}

\insertFigure{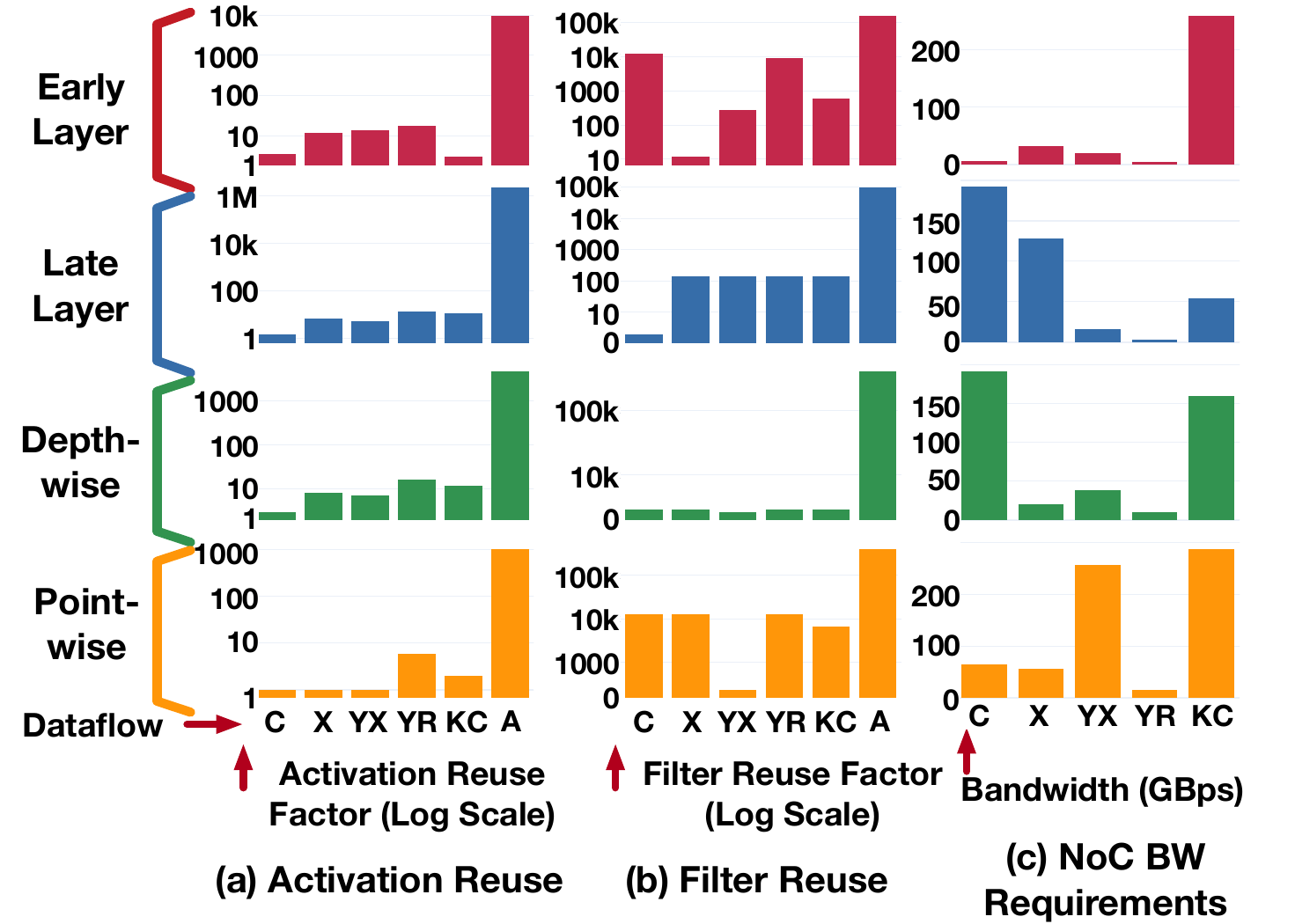}{Reuse and NoC bandwidth requirements of dataflows in~\autoref{table:EvalDataflows} with 256 PEs for four common DNN operators from~\autoref{table:DNN_Operators}. We select representative operators from state-of-the-art DNN models (Early layer: CONV1 in Resnet50~\cite{Resnet}, late layer: CONV13 in VGG16~\cite{VGGnet}, depth-wise convolution (DWCONV): DWCONV of CONV2 in ResNeXt50~\cite{ResNeXt}, point-wise convolution: first conv of bottleneck1 in MobilenetV2~\cite{mobilenetv2} C, X, YX, YR, and KC refers to C-P, X-P, YX-P, YR-P, and KC-P dataflows. A refers to algorithmic maximum reuse.).
}

\insertFigure{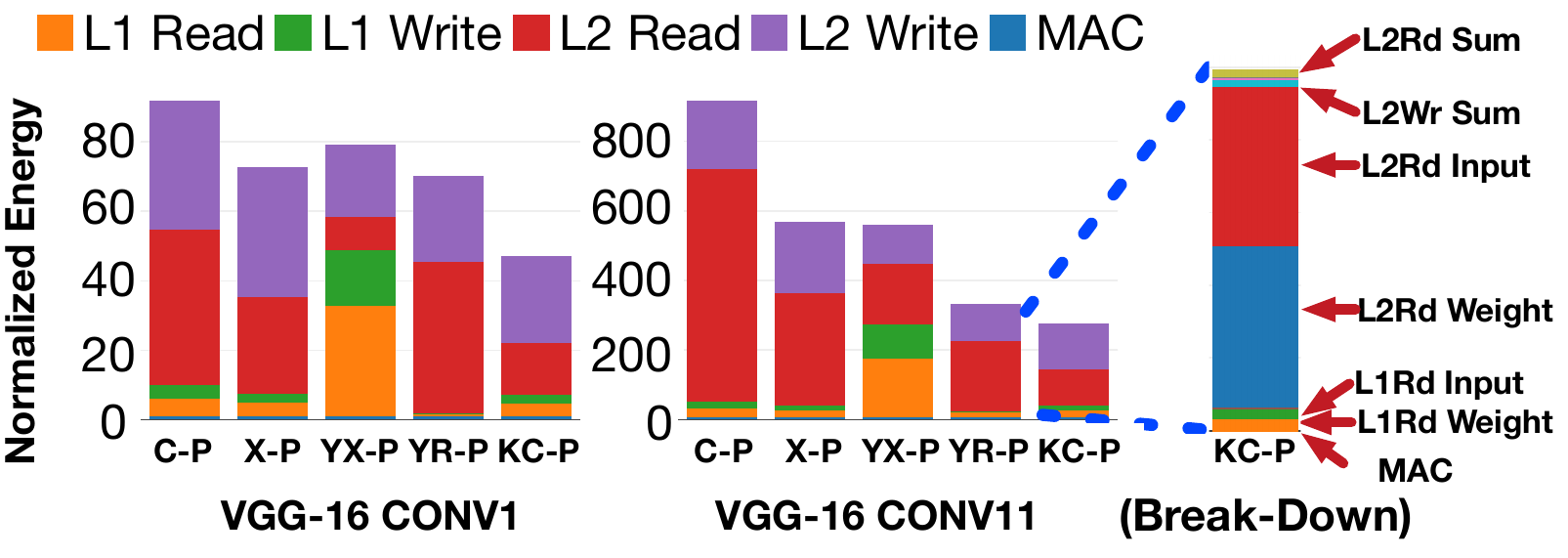}{The breakdown of energy consumption (MAC and L1/L2 scratchpad access energy) of the dataflows from~\autoref{table:EvalDataflows}. The access counts generated by \tool are multiplied by appropriate energy values from Cacti~\cite{muralimanohar2009cacti}. The values are normalized to the MAC energy of C-P.}



~\autoref{fig:RuntimeEnergy_LayerType} shows the DNN-operator granularity estimation of runtime and energy of each dataflow across five state-of-the-art DNN models listed in~\autoref{sec:eval}. Note that this should be considered a comparison of dataflows---not of actual designs, which can contain several low-level implementation differences, e.g.,
custom implementations of logic/memory blocks,
process technology, and so on.
We observe that KC-P dataflow style dataflow provides overall low runtime and energy.
However, the energy efficiency in VGG16 (\autoref{fig:RuntimeEnergy_LayerType} (b)) is worse than YR-P (Eyeriss~\cite{eyeriss_isca} style) dataflow, and the runtime is worse than YX-P (Shidiannao~\cite{du2015shidiannao} style) dataflow in UNet (~\autoref{fig:RuntimeEnergy_LayerType} (e)).
This is based on the different preference toward dataflow of each DNN operator.
YX-P provides short runtime to segmentation networks like UNet, which has wide activation (e.g., 572x572 in the input layer) and recovers the original activation dimension at the end via up-scale convolution (e.g., transposed convolutions).
Such a preference to the YX-P style is mainly based on its parallelization strategy: it exploits parallelism over both of row and column dimensions in activation.
%
%
The energy efficiency of YR-P dataflow in VGG16 is based on its high reuse factor (the number of local accesses per fetch) in early layers, as shown in red bars in~\autoref{fig:Dataflow_Analysis_Large_rev} (a) and (b) (note the log scale).
The YR-P dataflow has 5.8$\times$ and 15.17$\times$ higher activation and filter reuse factors, respectively, in early layers.
However, in late layers, the reuse factors of YR-P and KC-P dataflow are almost similar (difference < 11\%), so the KC-P dataflow provides similar energy efficiency as YR-P in these cases. 
This can also be observed in the late layer (blue) bars in~\autoref{fig:RuntimeEnergy_LayerType} bottom-row plots. 

Although the KC-P and YX-P dataflows provide low runtime (\autoref{fig:RuntimeEnergy_LayerType}), it comes with high NoC cost, as the high bandwidth requirements shown in~\autoref{fig:Dataflow_Analysis_Large_rev} (c) highlight. 
Based on the operator type, some dataflows require dramatically higher NoC bandwidth than others.
For example, YX-P requires high bandwidth for point-wise convolution as it has no convolutional reuse (i.e., overlapped activation data points among sliding windows) because of its 1x1 kernel while YX-P is optimized to exploit convolutional reuse via spatial reuse.

The diverse preference to dataflows of different DNN operators motivates us to employ optimal dataflow for each DNN operator type.
We refer such an approach as adaptive dataflow and present the benefits in~\autoref{fig:RuntimeEnergy_LayerType} (f), the average case analysis across entire models in DNN operator granularity.
By employing the adaptive approach, we could observe a potential 37\% runtime and 10\% energy reduction.
Such an optimization opportunity can be exploited by flexible accelerators like Flexflow~\cite{lu2017flexflow} and MAERI~\cite{kwon2018maeri} or via heterogeneous accelerators that employ multiple sub-accelerators with various dataflow styles in a single DNN accelerator chip.

\subsection{Case study II: Hardware Design-Parameters and Implementation Analysis}
\label{subsec:dse_eval}


\insertWideFigure{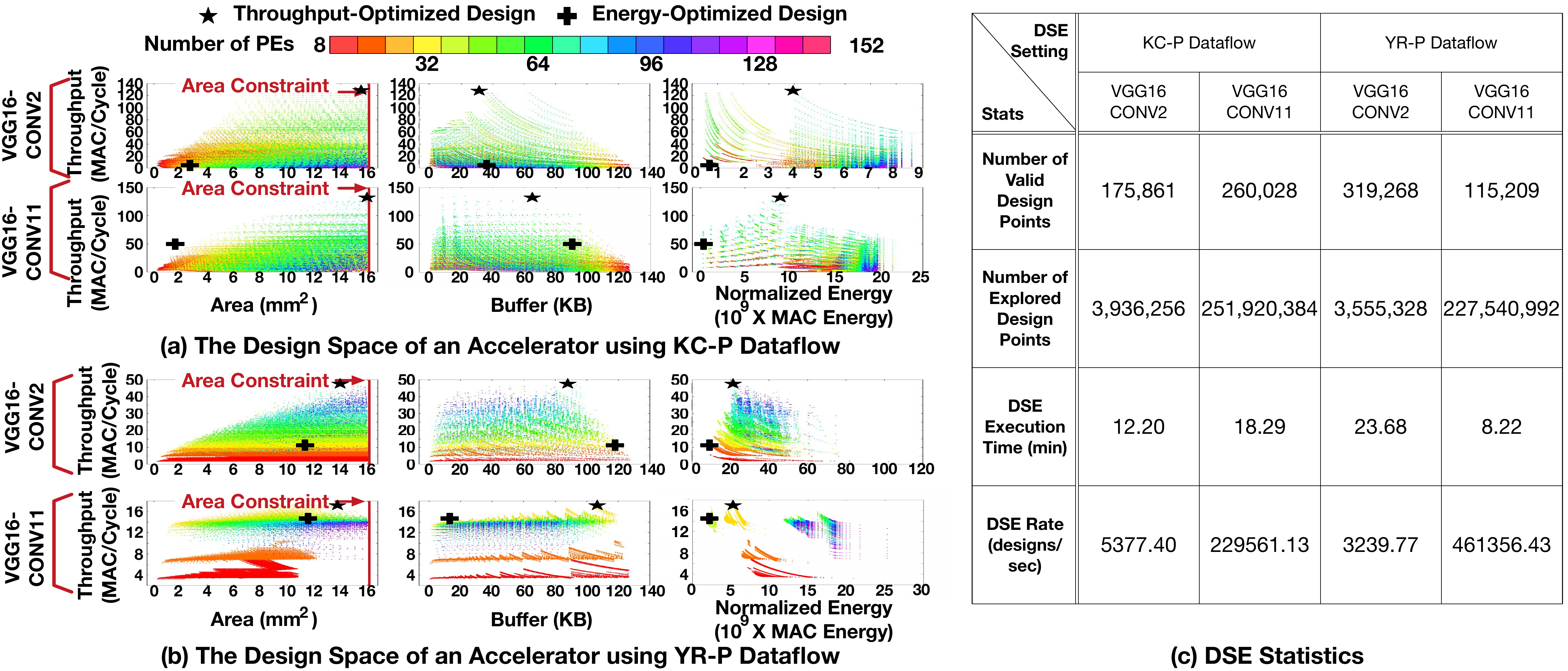}{The design space of an accelerator with (a) KC-P and (b) YR-P dataflow. We highlight the design space of an early and a late layer to show their significantly different hardware preference. We apply the area and power of Eyeriss~\cite{chen2017eyeriss_issc} as area/power constraints to the DSE.(16mm$^2$, 450mW~\cite{chen2017eyeriss_issc}). The color of each data point indicates the number of PEs. Design points with fewer PEs can be paired with larger buffer sizes, up to the area budget. We mark the throughput- and energy-optimized designs using a star and cross. } 



\insertWideTableFigure{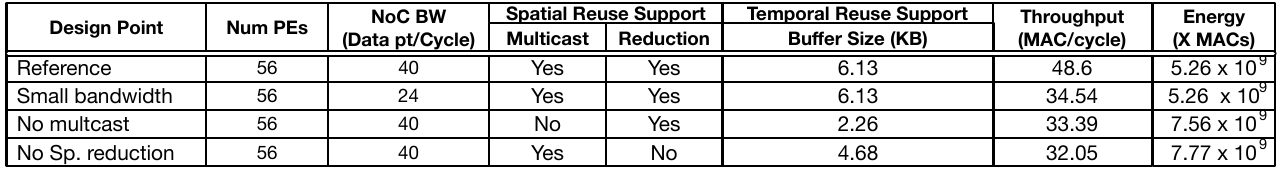}{The impact of multicasting capability, bandwidth, and buffer size. Design points are from the design space of ~\autoref{fig:DSE_ALL_HALF} (a) VGG16-CONV2.}


Using \tool, we implement a hardware design space exploration (DSE) tool that searches four hardware parameters (the number of PEs, L1 buffer size, L2 buffer size, and NoC bandwidth) optimized for either energy efficiency, throughput, or energy-delay-product (EDP) within given hardware area and power constraints.
The DSE tool receives the same set of inputs as \tool with hardware area/power constraints and the area/power of building blocks synthesized with the target technology.
For the cost of building blocks, we implement float/fixed point multiplier and adder, bus, bus arbiter, and global/local scratchpad in RTL and synthesis them using 28nm technology. For bus and arbiter cost, we fit the costs into a linear and quadratic model using regression because bus cost increases linearly and arbiter cost increases quadratically (e.g., matrix arbiter).
%
%

The DSE tool sweeps a target design space specified in the range of each parameter and search granularity.
However, it skips design spaces at each iteration of hardware parameters by checking the minimum area and power of all the possible design points from inner loops of hardware parameters.
This optimization allows it to skip invalid design points in a various granularity that reduces a large number of futile searches, which led to a large effective DSE rate ranging from 3.3K to 0.46M designs per second, as presented in~\autoref{fig:DSE_ALL_HALF} (c).
~\autoref{fig:DSE_ALL_HALF} (c) shows statistics of four DSE runs explored the design space. We ran DSEs on a machine with i7-8700k CPU and 32GB memory operating Linux Mint 19 OS. We run four sets of the DSE on the machine at the same time, and all of them terminated within 24 minutes, with effective DSE rate of 0.17M designs per second, on average.
\colonparagraph{Design Space Analysis} 
Using the DSE tool, we explore the design space of KC-P and YR-P dataflow accelerators. We set the area and power constraint as 16mm$^2$ and 450mW, which is the reported chip area and power of Eyeriss~\cite{chen2017eyeriss_issc}. We plot the entire design space we explored in~\autoref{fig:DSE_ALL_HALF}.

Whether an accelerator can achieve peak throughput depends on not only the number of PEs but also NoC bandwidth. In particular, although an accelerator has sufficient number of PEs to exploit the maximum degree of parallelism a dataflow allows, if the NoC does not provide sufficient bandwidth, the accelerator suffers a communication bottleneck in the NoC. Such design points can be observed in the area-throughput plot in ~\autoref{fig:DSE_ALL_HALF} (a). YR-P dataflow requires low NoC bandwidth as shown in~\autoref{fig:Dataflow_Analysis_Large_rev} (c) so it does not show the same behavior as KC-P dataflow.
However, with more stringent area and power constraints, YR-P dataflow will show the same behavior.

During DSE runs, \tool reports buffer requirements for each dataflow and the DSE tool places the exact amount buffers \tool reported. Contrary to intuition, larger buffer sizes do not always provide high throughput, as shown in buffer-throughput plots in~\autoref{fig:DSE_ALL_HALF} (plots in the second column). The optimal points regarding the throughput per buffer size are in the top-left region of the buffer-throughput plots. The existence of such points indicates that the tiling strategy of the dataflow (mapping sizes in our directive representation) significantly affects the efficiency of buffer use.

We also observe the impact of hardware support for each data reuse, discussed in~\autoref{table:HW_Impl_choices}.
~\autoref{table:DesignTweaks} shows such design points found in the design space of KC-P dataflow on VGG16-conv2 layer presented in the first row of ~\autoref{fig:DSE_ALL_HALF} (a). The first design point is the throughput-optimized design represented as a star in the first row of ~\autoref{fig:DSE_ALL_HALF}. 
When bandwidth gets smaller, the throughput significantly drops, but energy remains similar.
However, the lack of spatial multicast or reduction support resulted in approximately 47\% energy increase, as the third and fourth design points shows.

We observe that the throughput-optimized designs have a moderate number of PEs and buffer sizes, implying that hardware resources need to be distributed not only to PEs but also to NoC and buffers for high PE utilization. Likewise, we observe that the buffer amount does not directly increase throughput and energy efficiency. These results imply that all the components are intertwined, and they need to be well-balanced to obtain a highly-efficient accelerator.


\section{Related Works}
\label{sec:related}

\colonparagraph{Hardware DSE and dataflow optimization}
Dataflow optimization is one of the key optimization targets in many recent DNN accelerators such as Eyeriss~\cite{eyeriss_isca}, Flexflow~\cite{lu2017flexflow}, SCNN~\cite{parashar2017scnn}, and NVDLA~\cite{nvdla}.
C-brain~\cite{song2016cbrain}, Flexflow~\cite{lu2017flexflow}, and analyzed the cost-benefit tradeoff of three dataflows and explored the opportunity of adaptive dataflow based on the tradeoff.
Ma et al.~\cite{ma2017optimizing} also constructed an analytic model for convolutions on FPGAs focusing on three loop transformations; interchange, unroll, and tiling.
Although their analytic model provides an intuition for trade-offs of dataflows, the model focus on one dataflow style they propose, does not consider regional spatial reuse, spatio-temporal reuse opportunities in DNN accelerators, and also doesn't consider communication delay in the NoC, which can dominate for dataflows with large tile sizes.
Also, the target dataflow is optimized for HLS flow, and it requires expressing using complex annotated loop nest with HLS synthesis directives.
%
Caffeine~\cite{zhang2018caffeine} proposed a full automatic FPGA flow that includes pragma-based annotation in programs, dataflow optimization framework, and DSE for FPGAs based on the analytic model defined over loop tiling and unrolling.
However, the dataflow search space is limited due to fixed loop orders; three presets termed straightforward, input-major, and weight-mapping.

{\bf Past works related to data-centric approaches}: There have been some works related to exploring data-centric approaches~\cite{Kodukula2001, Kodukula:1997:DMB:258915.258946,Kodukula:1999:EET:305138.305243}, where the approaches reason about flow of data through memory hierarchy, instead of control-structure centric analysis, for locality-enhancement transformations such as multi-level data blocking ~\cite{Kodukula2001} and data shackling~\cite{Kodukula:1997:DMB:258915.258946}.
But, the above data-centric approaches have been explored in the context of driving optimizations for multi-level caches, but not estimating energy or throughput of input kernels precisely.
We discuss related work on loop-nest notation and reuse analysis in compilers in  ~\autoref{subsec:dataflowdescription}.

\section{Discussion and Future work}
\label{sec:conclusion}

This work is motivated by the observation that co-optimizing DNN accelerator microarchitecture and its internal dataflow(s) is crucial for accelerator designers to achieve both higher performance and energy efficiency. 
%
%
In this work, we introduced data-centric directives to specify DNN dataflows in a compact form and understand data reuse opportunities. We also presented an analytical model called \tool to estimate execution time, energy efficiency, and the hardware cost of dataflows.
We evaluated our analytical model relative to the MAERI and Eyeriss accelerators and found our model to be highly consistent with cycle-accurate simulations and reported runtime, which shows the soundness of the analytic model.
We provided cases studies about the costs and benefits of dataflow choices over in five state-of-the-art DNN models with a focus on common DNN operators in them, showing diverse preference to dataflow and hardware, which motivates adaptive dataflow accelerator and heterogeneous accelerators.
Finally, we also demonstrated the use of MAESTRO for design-space exploration of two dataflows in early and late layers, showing dramatically different hardware preference of each layer.
Our DSE tool based on \tool enabled fast DSE based on optimization to skip invalid designs, which led to a high average DSE rate of 0.17M designs per second.

In the future, we plan to leverage \tool to implement a dataflow auto-tuner to find an optimal dataflow on the specified DNN model and hardware configuration. 
With the optimal dataflow, we plan to extend our infrastructure to automatically generate RTL, facilitating end-to-end DNN acceleration flow.

\section*{Acknowledgement}

We thank Joel Emer for insightful advice and constructive comments to improve this work; Vivienne Sze and Yu-Hsin Chen for their insights and taxonomy that motivated this work.
This work was supported by NSF Awards 1755876 and 1909900.


\bibliographystyle{ACM-Reference-Format}
\bibliography{refs}

\end{document}